\documentclass[referee,useAMS,usenatbib]{mn2e}
\usepackage{graphicx}
\usepackage{rotating}
\usepackage{lscape}
\usepackage{deluxetable}
\usepackage{longtable}

\title[INOV of TeV blazars]
  {Rapid optical variability of TeV blazars}
\author[Gopal-Krishna et al.]
  {Gopal-Krishna$^1$\thanks{E-mail: krishna@ncra.tifr.res.in},
   Arti Goyal$^{1,2}$,
   S. Joshi$^{2}$,
   Chrisphin Karthick$^{2}$,
   R. Sagar$^{2}$,
\newauthor
   Paul J. Wiita$^{3}$,
   G. C. Anupama$^{4}$,
   D. K. Sahu$^{4}$\\
$^1$ National Centre for Radio Astrophysics/TIFR, Pune University Campus, Pune 411 007, India\\
$^2$ Aryabhatta Research Institute of Observational Sciences (ARIES),
Manora Peak, Naini Tal 263 129, India\\
$^3$ Department of Physics, The College of New Jersey, P.O. Box 7718, Ewing, NJ 08628, USA \\
$^4$ Indian Institute of Astrophysics (IIA), Bangalore 560 034, India \\
}

\date{Released 2002 Xxxxx XX}

\pagerange{\pageref{firstpage}--\pageref{lastpage}} \pubyear{2007}

\def\LaTeX{L\kern-.36em\raise.3ex\hbox{a}\kern-.15em
    T\kern-.1667em\lower.7ex\hbox{E}\kern-.125emX}

\begin{document}

\label{firstpage}

\maketitle

\begin{abstract}

In this first systematic attempt to characterise the intranight optical
variability (INOV) of TeV detected blazars, we have monitored a well
defined set of 9 TeV blazars on total 26 nights during 2004-2010.
In this $\it R$ (or $\it V$)-band monitoring programme only one blazar 
was monitored per night and the minimum duration was close to 4 hours, the 
average being 5.3 hours per night. 
Using the CCD for strictly simultaneous photometry of the blazar and 
nearby reference stars (N-star photometry), 
an INOV detection threshold of $\sim$ 1--2 per cent was achieved in the 
densely sampled differential light curves derived from our data. We have 
further expanded the sample by including another 13 TeV blazars, taking 
advantage of the availability in the literature of INOV data, including those
published earlier in our programme. The selection criteria 
for this set of 13 blazars conform to the basic criteria we had adopted for 
the first set of 9 blazars we have monitored presently. 
This enlarged, well defined representative sample of 22 TeV blazars, monitored 
on a total of 116 nights (including 55 nights newly reported here),
has enabled us to arrive at the first estimate of 
the INOV duty cycle (DC) of TeV detected blazars. Applying the conservative, but 
commonly employed, $C$-test, the INOV DC is found to be 59 per cent, which decreases to
47 per cent if only INOV fractional amplitudes ($\psi$) above 3 per cent are considered. 
These observations
also permit, for the first time, a comparison of the INOV characteristics 
of the two major subclasses of TeV detected BL Lacs, namely LBLs and HBLs, for which
we find the INOV duty cycles to be $\sim$ 63 per cent and $\sim$ 38 per cent,
respectively. This 
demonstrates that the previously recognized INOV differential between 
LBLs and HBLs persists even when only their TeV detected subsets are considered.
Despite dense sampling, the intranight light curves of the 22 TeV blazars have 
not revealed even a
single feature on time scale substantially shorter than 1 hour, even though
the inner jets of TeV blazars are believed to have exceptionally large bulk
Lorentz factors (and correspondingly stronger time compression). 
An intriguing feature, clearly detected in the light curve 
of the HBL J1555$+$1111, is a 4 per cent `dip' on a 1 hour timescale. This 
unique feature 
could have arisen from absorption in a dusty gas cloud, occulting a superluminally 
moving optical knot in the parsec scale jet of this relatively luminous BL Lacs object.

\end{abstract}


\section{Introduction}

Intensity variations have long been recognized as a defining characteristic
of active galactic nuclei (AGN).  Variability is a powerful tool for probing AGN 
geometry and physical properties such as the black hole mass, and the sizes and bulk 
motions of the outflows in their innermost regions that are well beyond the 
current imaging capabilities of telescopes in any part of the electromagnetic 
spectrum (e.g.,  Wagner \& Witzel 1995; Urry \& Padovani 
1995; Xie et al.\ 2001). The variations can be particularly violent for those 
AGNs whose flux is dominated by relativistic jets of nonthermal radiation
broadly pointing in our direction (e.g., Begelman, Blandford \& Rees 1984).
Intensities of such AGNs, called `blazars', are known to vary across the entire
electromagnetic spectrum and time scales from minutes to years have been 
observed. For instance, in the soft X-ray band, several TeV emitting blazars 
have been found to vary on a characteristic time scale of $\sim$1 day, with 
the flares having sub-structures on shorter time scales of  $\sim10^4$s 
(e.g., Tanihata et al.\ 2001; Kataoka et al.\ 2001). In the optical 
regime, there have been many detections of intra-night optical variability 
(INOV), or optical microvariability, following the pioneering work of Carini, 
Miller \& Goodrich (1990) who first used CCD detectors as multi-object
photometers for this purpose. The shortest time scale found for essentially all INOV 
events is around 1 hour, with an amplitude of a few percent (e.g., Xie et 
al.\ 2001; Romero et al.\ 2002; Stalin et al.\ 2004a). Such rapid continuum 
variability of blazars is usually explained by invoking relativistic jets 
(e.g., Marscher 1996; Schlickeiser 1996; Wiita 2006).

A study of 23 AGN of the quasar/Seyfert1 type (i.e., non-blazar) yielded a 
$1 \sigma$ upper limit of 0.03 mag for variability on hour-like timescales 
(Webb \& Malkan 2000). The literature does contain reports of a few INOV 
detections on time scales much shorter than 1 hour. Examples include the papers 
by Kidger \& deDiego (1990), 
Sagar, Gopal-Krishna \& Wiita (1996), Xie et al.\ (2001), and Dai et al.\ (2001); however, see 
Romero et al.\ (2002) for a convincing critique of the latter two claims. To our 
knowledge such assertions are lacking in the literature over the past 5 years 
or so, aside from a very recent paper presenting evidence for a quasi-periodic 
oscillation of $\simeq $ 15 min spanning most of a night of optical monitoring 
of the TeV blazar S5 0716$+$714 (Rani et al.\ 2010), which is also a member of 
our present sample. Very recently, Impiombato et al. (2011) have reported a single event 
of duration $\sim$25 minutes in $J$-band while searching for long and short term 
optical and infrared variability for blazar PKS0537$-$441 in the data spanning 
for $\sim$6 years (2004-2009).
There is thus a need for renewed efforts capable of making 
robust detections of events of ultra-rapid INOV events on sub-hour time scales.

Since  $\gamma-$ray emission is known to be correlated with relatively 
large bulk Doppler factors of the radiating plasma in the blazar jets
for both the GeV band probed by EGRET (e.g., Kellermann et al. 2004; Lister \& Homan 2005)
and the higher energies probed by Fermi/LAT (Kovalev et al.\ 2009; Savolainen et al.\ 2010), 
TeV blazars appear to be 
particularly promising candidates for detecting the most rapid INOV. In TeV
blazars, the relativistic plasma of the inner jets almost certainly must 
move with a bulk Lorentz factor $\Gamma \geq$ 45--50 in order to escape 
absorption of the TeV photons due to the pair production process in the 
radiation field present near the origin (e.g., 
Krawczynski, Coppi \& Aharonian 2002; Begelman, Fabian \& Rees 2008; for 
large Lorentz factors in blazar jets, also see Kundt \& Gopal-Krishna 1980,
2004). According to one model, such ultra-fast moving emission features may 
arise {in situ} within the jet, e.g., from reconnection events in a Poynting 
flux dominated jet (Giannios, Uzdensky \& Begelman 2009). 
Thus, it seems likely that in TeV blazars the bulk Lorentz factors of the 
optically radiating inner segments of the jets are also comparatively 
larger than those occuring in other blazars. Here we note that the marked 
paucity of apparent superluminal motion in the jets of TeV blazars, 
highlighted by Piner \& Edwards (2004), can be reconciled with the above 
requirement of extreme bulk Lorentz factors in a number of ways, e.g., by 
taking into account a modest opening angle for the jet (Gopal-Krishna, Dhurde 
\& Wiita 2004; Gopal-Krishna, Wiita \& Dhurde  2006; Gopal-Krishna et al.\ 
2007; Petrucci, Boutelier \& Henri 2010), or by postulating a spine-sheath 
geometry for the sub-parsec scale jets (e.g., Attridge, Roberts \& Wardle 1999; 
Ghisellini, Haardt \& Matt 2004) or, alternatively, if the bulk of the $\gamma-$rays
come from an ultra-relativistically approaching volume element in the jet (e.g.,
Giannios et al.\ 2009; also, Gopal-Krishna, Singal \& Krishnamohan 1984).

In Sect.\ 2 we  describe the selection of our sample of TeV blazars 
for intranight optical monitoring. Our observations are described in Sect.\ 3 
and the results summarized in Sect.\ 4, followed by a discussion in Sect.\ 5.

\section{Sample selection for TeV blazars}

Our sample of TeV blazars consists of two sets. Set 1 is derived from the 
list of TeV detected extragalactic AGN, published by Weekes (2008; Table 2 of his paper). 
Application of the criteria, $z > 0.1$, $\delta > -10^{\circ}$ and $m_B 
\le18$ to that list leaves us with 9 TeV blazars. The redshift limit is 
designed to exclude the nearest blazars so that the optical image should have
a point-like appearence. This is to ensure that the host galaxy makes a 
negligible contribution to the image, a pre-requisite for high precision 
photometry. The declination and the apparent magnitude limits are meant to
ensure a sufficiently dense sampling and reasonably long duration ($\ga 4$ hours) 
for the continuum light curves obtained with 
the 1 $-$ 2 metre class telescopes employed in this programme. The 9 TeV blazars 
constituting Set 1 are: TeV1219$+$283, TeV1429$+$427, TeV0809$+$524, 
TeV1218$+$304, TeV1011$+$496, TeV0716$+$714, TeV1553$+$11, TeV0219$+$425
and TeV1256$-$058, listed in increasing order of distance from us. The intra-night 
and long-term optical lightcurves of these 9 blazars are derived in
this study and presented here.

Our Set 2 of TeV blazars was derived from Table 1 of Abdo et al.\ (2010),
consisting of 709 TeV detected AGNs, which is based on 11 months of monitoring with 
$\rm Fermi$ $ \rm LAT$. This set consists of 13 blazars, including 10 high 
polarization core-dominated quasars (HPCDQs) and 3 BL Lac objects. The HPCDQs 
were selected employing the following criteria:  
all CDQs labeled ``HP'' ($P_{op} > 3$ per cent) in the compendium of V\'eron \& 
V\'eron (2006) were subjected to the aformentioned selection criteria, namely,
$z > 0.1$, $\delta > -10^{\circ}$ and $m_B \le18$.
To ensure the availability of INOV data additional selection filters used are:

(a) We selected all the 7 HPCDQs from J. Noble's PhD thesis (Noble 1995, 
    Table 3.1). 
    These are J0239$+$1637, J1159$+$2914, J1256$-$0547,
    J1310$+$3220, J1643$+$3953, J2225$-$0457, J2253$+$1608.
    Of these J1256$-$0547 (3C 279) is already a member of Set 1 and J1643$+$3953 
    (3C 345) is undetected by $\rm Fermi$ $\rm LAT$. This gave us 
    the first 5 TeV blazars for Set 2.

(b) Another 3 HPCDQs were taken from the polarimetric survey of Wills et al. 
    (1992), limiting ourselves to the right ascension range $03^h$ -- $15^h$
    and declination range $-$$10^\circ$ to +40$^\circ$. Note that although 
    these criteria yielded 5 HPCDQs (J0423$-$0120, J0739$+$0136, J1058$+$0133, 
    J1159$+$2914 and J1256$-$0547), the last two of these are already a part 
    of the Noble (1995) sample (a). This left us with 3 additional  blazars. 

(c) One HPCDQ, J1218$-$0119, was taken from the first phase of our INOV
    program (Sagar et al.\ 2004, Stalin et al.\ 2005). INOV data have been 
    taken from those papers for this source as well as for another two HPCDQs
    (J0239$+$1637 and J1310$+$3220) that are in the part (a) of Set 2 derived 
    from Noble (1995). Note that these were the only 3 HPCDQs monitored in 
    the first phase of our INOV program. 

(d) Lastly, one HPCDQ was added from the sample of Romero, Cellone \& Combi 
    (1999). They have reported V-band intranight monitoring of a sample of 
    southern AGN that contains 4 HPCDQs, according to the V\'eron \& V\'eron 
    (2006) classification. These are J0538$-$4405, J1147$-$3812, J1246$-$2547,
    and J1512$-$0906. Only the last of these HPCDQs was included in our sample; 
   because we could ensure a minimum 3 nights' monitoring data;
    the others are situated far to the south. 
   
(e) We have also included the 3 BL Lacs PKS 0735$+$178, OJ 287 and B2 1215$+$30 
    for which intra-night monitoring data of duration $\ga$ 4 hours are 
    available in the literature. This completes our Set 2 that contains 13 
    TeV detected blazars.

Salient properties of the two blazar sets are listed in Table 1.  All these sources
have flat radio spectra ($\alpha_r > -0.5$, where 
$S(\nu) \propto \nu^{\alpha_r}$) as well as high 
optical linear polarization, with $P_{op}$ falling in the range 3.5 to 44 per 
cent, except for J1428$+$4240 (which has a slightly steep integrated radio spectrum with 
$\alpha_{r} \simeq -0.58$, and a 
comparatively low optical polarization, $P_{op}$ = 2.5\%). 
The values of radio core luminosity ($P_c$), extended radio 
luminosity ($P_{ext}$), and the radio core-dominance parameter ($f_c$; ratio 
of core to extended radio luminosities) at 5 GHz, were determined using 
the available Very Long Baseline Interferometry (VLBI) measurements 
at milli-arcsecond resolution and the integrated 
NRAO VLA Sky Survey (NVSS) flux values at 1.4 GHz, taking a radio spectral index of zero for the 
core ($\alpha_{c} =0$) and $\alpha_{ext} = -0.5$ for the extended emission.
The concordance cosmological model was assumed, with a Hubble constant $H_{0} 
= 70$ km sec$^{-1}$ Mpc$^{-1}$, $\Omega_{m} =0.3$ and $\Omega_{\Lambda} =0.7$ 
(Spergel et al. 2007).
Values of the radio loudness parameter ($R^*$) were determined following
Stocke et al.\ (1992). The absolute magnitudes, $M_B$, were calculated 
taking the total galactic extinction from Schlegel, Finkbeiner \& Davis (1998) 
and assuming an optical spectral index $\alpha_{op}$ of $-0.7$.

\section{Observations}
\subsection{Instruments used}
The observations were mainly carried out using the 104-cm
Sampurnanand telescope (ST) located at the
Aryabhatta Research Institute of observational sciencES (ARIES), in
Naini Tal, India. It has Ritchey-Chr\'etien (RC) optics with
a f$/$13 beam (Sagar 1999). The detector was a cryogenically
cooled 2048 $\times$ 2048 chip mounted at the Cassegrain focus.
This chip has a readout noise of 5.3 e$^{-}$/pixel and a gain 
of 10 e$^{-}$$/$Analog to Digital Unit (ADU) in the usually 
employed slow readout mode. Each pixel has a dimension of 
24 $\mu$m$^{2}$ which corresponds to 0.37 arcsec$^{2}$ on 
the sky, covering a total field of 13$^{\prime}$ $\times$
13$^{\prime}$. Observations were carried out in 2 $\times$ 
2 binned mode to improve the S$/$N ratio. All the observations with the ST
were carried out using an {\it R} filter for which the CCD 
sensitivity is maximum. The seeing ranged mostly between
$\sim 1^{\prime\prime}.5$ and $\sim$3$^{\prime\prime}$, 
as determined using 3 moderately bright stars within the CCD frame.
For each night, the plot of seeing is provided in Figure 1 in the
corresponding bottom panel.

The other telescope used for our monitoring of TeV blazars is the 201-cm Himalayan Chandra 
Telescope (HCT) of the Indian Astronomical Observatory (IAO) located at 
Hanle, India, which is also of the RC design with a f$/$9 beam at the
Cassegrain focus\footnote{http://www.iiap.res.in/$\sim$iao}.
The detector was a cryogenically cooled 2048 $\times$ 4096 chip, 
of which the central 2048 $\times$ 2048 pixels were used. 
As the pixel size is 15 $\mu$m$^{2}$ the image scale of 
0.29 arcsec$/$pixel covers an area of 10$^{\prime}$ $\times$ 10${^\prime}$ 
on the sky. The readout noise of the CCD is 4.87 e$^{-}$/pixel and the 
gain is 1.22 e$^{-}$$/$ADU. This CCD was used in an unbinned mode. 
The seeing ranged mostly between $\sim1^{\prime\prime}$ to 
$\sim3^{\prime\prime}$.

Lastly, one night of blazar monitoring data were obtained using the 200-cm IUCAA Girawali
Observatory (IGO) telescope located near Pune, India. It has an RC design
with a f$/$10 beam at the Cassegrain focus\footnote
{http://www.iucaa.ernet.in/\%7Eitp/igoweb/igo$_{-}$tele$_{-}$and$_{-}$inst.htm}.
The detector was a cryogenically cooled 2110$\times$2048 chip mounted
at the Cassegrain focus. The pixel size is 15 $\mu$m$^{2}$ so that the
image scale of 0.27 arcsec$/$pixel covers an area of 10$^{\prime}$
$\times$ 10${^\prime}$ on the sky. The readout noise
of this CCD is 4.0 e$^{-}$/pixel and the gain is 1.5 e$^{-}$$/$ADU.
The CCD was used in an unbinned mode. The seeing ranged 
between $\sim$1$^{\prime\prime}$.0 and $\sim$1$^{\prime\prime}$.5 on that particular night.

The observations were made using {\it R} or {\it V} filters.
The exposure times were typically 10 to 12 minutes for the 
ARIES and IGO observations and ranged between 3 to 6 minutes for 
the observations from IAO (depending on the brightness of the source, 
the lunar phase and the sky transparency for the night). 
The field positioning was adjusted so as to always have within the
CCD frame 2--3 comparison stars within about a magnitude of the blazar, 
in order to minimize the possibility of getting spurious 
variability detection (e.g., Cellone, Romero \& Araudo 2007). 
For all three telescopes,  bias frames were  taken
intermittently and twilight sky flats were obtained on each night.
Each blazar in Set 1 was monitored for a minimum of 3 nights. Likewise, for the
blazars in Set 2, this requirement of minimum 3 nights is met, except for
the case of J2253+1608 (3C 454.3) for which only 2 nights' of monitoring
data are available. 

\subsection{Data reduction}
The preprocessing of the images (bias subtraction, flat-fielding 
and cosmic-ray removal) was done by applying the standard
procedures in IRAF\footnote{\textsc {Image Reduction
and Analysis Facility (http://iraf.noao.edu/)}} and MIDAS\footnote{\textsc 
{Munich Image and Data Analysis System 
(http://www.eso.org/sci/data-processing/software/esomidas//)}} softwares. 
The instrumental magnitudes of the blazar and the comparison stars 
in the image frames were determined by aperture photometry 
using DAOPHOT \textrm{II}\footnote{\textsc {Dominion
Astrophysical Observatory Photometry} software} (Stetson 1987).
The magnitude of the blazar was measured relative to the 
steady comparison stars present on the same CCD frame 
(Table 2). In this manner, Differential Light Curves (DLCs) of 
each blazar were produced relative to 3 comparison stars
(Fig. 1).
For each night, the selection of optimum aperture size for the photometry
was done by examining the observed dispersions in the 
star-star DLCs for different aperture radii starting from 
the median seeing (FWHM) value on that night to 4 times 
that value.  We selected the aperture that showed minimum scatter for the steadiest DLC found for 
the various pairs of the comparison stars (e.g., Stalin et al.\ 2004a). 

\section{Results}

\subsection{Variability criteria and duty cycles}

The INOV DLCs are shown in Figure 1 for all the 9 blazars of Set 1. From
Set 2 we present here the DLCs for just J0854$+$2006 (OJ 287) since it was also monitored
in the present work (Fig. 1). The DLCs for another few members of the Set 2, which
too were monitored in the present work, can be found in Goyal et al. (2011)
as they form part of the samples discussed in that paper.
Figure 2 displays the long-term optical variability (LTOV) DLCs for sources 
that could be observed in the same colour filter on a minimum of 3 nights.
Table 3 summarizes the observations and the derived results 
for our entire sample of 22 TeV blazars, 9 of which were monitored in
the present study (Set 1), while for the remainder (Set 2) the INOV 
data were largely taken from the literature (Sect.\ 2; Table 3).
The 6th, 7th and 8th columns give, respectively, the monitoring duration 
on the respective night, the number of data points (N$_{points}$) in the DLC, and the rms of the 
steadiest star--star DLC obtained using two of the comparison stars.

The next columns in Table 3 contain the measures of the source variability (INOV)
on each night.
The fractional amplitude of INOV, $\psi$, is given in Col.\ (9), while  $C_{\rm eff}$, a
widely used indicator of variability status, is in Col.\ (10). 
The classification `variable' (V) or `non-variable' 
(N) as decided using the  $C$-test, basically defined following the criteria of
Jang \& Miller (1997), is in Col.\ (11).  

$C_{\rm eff}$ for a night is derived by combining the $C$-values 
estimated for individual DLCs of the blazar on that night.  For a given DLC, $C$ is defined 
as the ratio of its standard deviation, $\sigma$$_T$ 
and $\eta\sigma_{err}$, where $\sigma_{err}$ is the 
average of the rms errors of its individual 
data points. As the photometric errors given by the  
$\rm DAOPHOT/IRAF$ package is known to be underestimated, 
a compasatory factor $\eta$ is determined that would make the 
rms of the DLC consistent with the rms of its individual data points
(see Edelson et al. 2002, also, Stalin et al. 2004a).
In this way, the computed value of $\eta$ is found to be $\sim$1.5 
(Stalin et al.\ 2004a, 2004b, 2005; Gopal-Krishna 
et al.\ 2003; Sagar et al.\ 2004). However, our analysis 
for the present dataset yields $\eta = 1.3$ and so we
have adopted this value here. We computed $C_{\rm eff}$
from the $C$ values (as defined above) derived for the 
individual DLCs of a given blazar relative to 3 different comparison stars 
which were monitored simultaneously with the blazar on the same CCD chip. 
This gave us 3 values of $C$ for a given blazar. 
These values were converted into probabilities and multiplied to 
obtain $C_{\rm eff}$ for the blazar (Sagar et al.\ 2004 for details) .  
This has the advantage of using 
the available multiple DLCs of an AGN, relative to different 
comparison stars. The AGN is termed `V' for 
$C_{\rm eff} >  2.576$, corresponding to a nominal confidence
level above 0.99. The `probable `variable' (PV) classification applies when, 
$1.950 < C_{\rm eff} < 2.576$, corresponding to a nominal confidence
level between 0.95 to 0.99. 

It has been recently argued by de Diego (2010), however, that 
$C$ is not a proper statistic as it is based on ratios of standard deviations
and not on ratios of variances; only the latter are distributed in such a way that
$\chi^2$ tests can be used to assign proper confidence levels. He shows
that the standard $F$-statistic, where
\begin{equation}
F =  \frac{\sigma^2(blazar-star_i)}{ \sigma^2(star_i-star_j)},
\end{equation}
is a more appropriate choice for characterizing AGN light curves.  It also
has the advantage that the somewhat uncertain parameter, $\eta$, cancels out.
Col.\ 12 gives the $F$-value for each night along with the number of degrees of 
freedom shown in parentheses. 
Since we can compute the $F$-statistic only for our data,
for the bulk of blazars in Set 2 (not observed by us) 
only the $C$-values are available.
In computing the $F$-values we examined the
various star--star light curves to decide if any of the comparison stars might
have even slightly varied on that night. We considered the steadiest two of the
three comparison stars and selected the one closer to the blazar in apparent
magnitude. The two DLCs (namely, blazar$-$star and star$-$star)
involving the selected comparison star were then used in the $F$-test (Eq. 1).
For a fair comparison to the $C$-test, we take for the $F$-test a 
significance of $>$ 0.99 to correspond to a definitely
variable source (V) and a significance between 0.95 and 0.99 to correspond to 
the PV classification. These indicators of variability status as computed using 
$F$-test are tabulated in the 13th column. The last column gives the reference(s) 
for the INOV data used here.

The peak-to-peak 
INOV amplitude is calculated using the definition 
(Romero et al., 1999)
\begin{equation}
\psi= \sqrt{({D_{max}}-{D_{min}})^2-2\sigma^2}
\end{equation}
with 
$D_{max}$ = maximum in the AGN's differential light curve,
$D_{min}$ = minimum in the AGN's differential light curve, and
$\sigma^2$= $\eta^2$$\langle\sigma^2_{err}\rangle$.

The INOV duty cycle (DC) for our entire sample of 22 TeV blazars (Table 3) 
was then computed following the definition of Romero et al. (1999) 
(see also, Stalin et al.\ 2004a):
\begin{equation}
DC  = 100\frac{\sum_{i=1}^n N_i(1/\Delta t_i)}{\sum_{i=1}^n (1/\Delta t_i)} {\rm per cent}
\label{eqno1}
\end{equation}
where $\Delta t_i = \Delta t_{i,obs}(1+z)^{-1}$ is the
duration of the blazar monitoring session on the $i^{th}$ night, corrected for
the blazar's cosmological redshift, $z$. Note that since the monitoring durations for
any given source on different nights were not equal,
the computation of DC has been weighted by the actual monitoring duration 
$\Delta t_i$ on the $i^{th}$ night. 
The parameter $N_i$ was set equal to 1 if INOV was detected; otherwise $N_i$ = 0.

We realize that 2 of our 9 blazars in Set 1 lie at rather small redshifts 
(Table 1; J1221+2813 at $z$ = 0.102 and J1428$+$4240 at $z = 0.129$), 
raising the possibility of a significant contribution from the host galaxy 
to the flux falling within the circular photometry aperture centered at the blazar. 
As stressed by, e.g., Cellone, Romero \& Combi (2000), the host's varying contribution 
to the flux within the aperture changes as the seeing 
varies and thus the atmospheric seeing changes can yield spurious INOV detection. 
The paper also presents simulated DLCs for AGN, relative to a suitable
comparison star, for cases where the emission from the AGN host galaxy
(elliptical or spiral) is comparable to that from the AGN itself and the
atmospheric seeing undergoes a large intranight variation. For a wide range
in the host galaxy size, they find that even if the host's flux is comparable
to that of the AGN (an extreme situation from the perspective of the present sample),
the DLCs show a negligible variation ($\psi < 1\%$) as long as the aperture
radius exceeds $\sim 3^{\prime\prime} - 4^{\prime\prime}$ and the seeing remains within this limit
(as also applicable to the present study). Therefore, since our choice of
aperture size always meets this condition, we do not expect any of our DLCs
classified as `V' to be spurious, i.e., being an artefact of variable
atmospheric seeing during the night.

Note also that in a very small number of cases the INOV results taken from the literature 
were for the V band; however, for the present purpose we do not distinguish them
from our data which were essentially always taken in the R band.

Based on the $C$-test the computed INOV DC for our sample of 
22 TeV blazars is found to be $\sim 59$ per cent (116 nights; Table 3),
which increases to $\sim 64$ per cent if five `PV' cases of `probable' INOV 
are also included. However, for the nights showing larger INOV amplitudes, 
$\psi > 0.03$, the INOV DC is $\sim 47$ per cent.  
 Since the redshift the blazar J1555+1111 is known to be controversial, 
with values ranging from 0.25 to 0.50 
(see, Treves, Falomo \& Uslenghi, 2007 and MAGIC collaboration: Albert et al. 2008)  
we have computed the $DC$ vlaues for our 
entire sample of blazars (116 nights), taking the lower and upper $z$ values for 
J1555+1111. The computed values are $DC$ = 58.6\% and $DC$ = 58.9\%, 
respectively. Thus, the uncertainty in $z$ of J1555+1111 does not significantly 
affect the estimated $DC$ for the sample (i.e., DC $\sim$59\%). 
Using the $F$-test, the corresponding $DC$ values for the 9 blazars monitored
presently, is $\sim 72$ per cent
($\sim 76 $ per cent, if the two `PV' cases are included) 
but only $\sim 33$ per cent for the cases having  $\psi > 0.03$ (55 nights; Table 3).

\subsection{Notes on the optical lightcurves of the TeV blazars} 

\subsubsection{\bf Set 1} 

The basic parameters for all these 9 blazars monitored by us have been taken 
from the compilations 
by Abdo et al.\ (2010a) and Weekes (2008) and their INOV and LTOV lightcurves 
determined in the present work are shown in Fig.\ 1 \&  Fig.\ 2, respectively.
As seen from Table 1, the Set 1 of TeV blazars consists of five HBLs, two IBLs, one 
LBL and one FSRQ.

\begin{itemize}
\item{{J0222$+$4302 (3C 66A; $z =0.444$)}

This blazar (Bramel et al.\ 2005) has been classified
as an intermediate-peaked BL Lac object (IBL) whose nonthermal emission peaks
in the range $10^{15}-10^{16}$ Hz (Perri et al. 2003). During 1998--2000 it was 
monitored in our programme on 7 nights, for durations ranging between 5 and 10 hours
per night. INOV was detected on 6 of the nights (Sagar et al.\ 2004). The
only sub-hour feature seen in those DLCs is a 1.5 per cent  `glitch' detected on the
night of 13 Nov.\ 1999 at 19.6 U.T. In Table 3 we have combined 
those published data with the two DLCs obtained in the present work. Of these, 
the DLC on 28 Sept.\ 2009 showed confirmed INOV using both $C_{\rm eff}$ and $F$ statistics.
Also, a clear event of {\it internight} variability was observed when the source 
faded by $\sim$0.04 mag between 27 and 28 Sept.\, 2009 (Fig. 1).}

\item{{J0721$+$7120 (S5 0716+714; $z =0.31$)}

This blazar with a `LBL' classification is the archetypal intra-day variable 
(e.g., Wagner et al.\ 1996) with a history of frequent 
high-amplitude flux variations (e.g., Gupta et al.\ 2008 and 
references therein). From their $R$-band light curves Wagner et al.\ (1996)
had reported a significant 
``flickering'' on time scale as short as $\approx 15$ minutes. 
Recently, its high temporal resolution study revealed quasi-periodic 
oscillations of about 15 minutes at $> 3 \sigma$ significance (Rani et al.\ 2010). 
The presence of such short time scales in an optical light curve can provide
important clues to relativistic beaming at these wavelengths 
(e.g., Fabian \& Rees 1979; Guilbert, Fabian \& Rees 1983).
During February and March 1994 this blazar was the target of a 4-week long
INOV monitoring campaign (BVRI) under our programme, using two Indian
telescopes: the 1-m ST and the 2-m Vainu Bappu Telescope  (Sagar et al. 1999). 
Ghisellini et al.\ (1997) have reported a multi-colour optical monitoring campaign
around that time, revealing a moderately active state of this blazar.
In the Indian campaign, a monitoring duration of 2--3 hours was typically 
achieved on each night. No evidence was found for INOV on time scale
shorter than 1 hr, but 3 prominent events on $\sim 3$ hour time scales
were detected. Also, {\it internight} variability with $\psi \sim 5 - 20$ per cent
was frequently detected during the 4-week long Indian campaign.
Results from other recent monitoring campaigns on this source performed at different 
sites should be available soon (J.\ Webb; A.\ Gupta, private communications).

The present one night's monitoring of this blazar detected a clear flare 
lasting $\approx 1$ hour (Fig. 1; Table 3). However, since the 1.5 hour
duration of our monitoring is much less than our selection 
criterion of $\ga$4 hour, we have not included this
observation in the computation of DC for this blazar (Table 3). Note that neither of
the aforementioned published observational campaigns revealed INOV time scale $\tau
< 1$ hour, even for this highly variable blazar, except for results reported by 
Wagner et al. (1996) and some recent observations reported by Rani et al.\ (2010).}

\item{{J0809$+$5218 (1ES 0806$+$524; $z =0.138$)}

To our knowledge, the present work is the first intranight optical monitoring 
of this HBL (4 nights, Table 3). A highly unusual aspect of this source, as revealed by the 
5 GHz VLBI observations, is a two-sided jet structure on parsec scale 
straddling the core (Chen et al.\ 2006), which itself has a brightness temperature of 
$\sim 10^{10}$ K, markedly lower than the inverse-Compton limit (Kellermann 
\& Paulini-Toth 1969). These unusual characteristics are, however, consistent 
with the absence of strong relativistic boosting in this HBL, as also reflected
by its quite stable GeV flux (Chen et al.\ 2006).  On none of the 4 nights 
did we detect INOV in this blazar (Table 3), although on longer term ($\sim 10$
month timescale), a clear fading by $\sim$0.45 mag was detected.  Thereafter, 
a distinct brightening by $\sim$ 0.04 mag occured between the observations 
taken 4 days apart (Fig.\ 3). }

\item{{J1015$+$4926 (1ES 1011$+$496; $z =0.212$)}

Out of the 3 nights observed by us this HBL showed a clear variability on 
one night, as confirmed by both $C_{\rm eff}$ and $F$-statistics (Table 3; 7 Mar.\ 2010).
On the night of 19 Feb 2010 it was found to be a probable variable using $C_{\rm eff}$ 
but classified as ``V'' on application of $F$-statistics. On a longer time scale,
 a steady fading by $\sim$0.11 mag was seen over the 1$-$month span covered by our monitoring 
(Fig.\ 3). }

\item{{J1221$+$3010 (1ES 1218$+$304; $z =0.182$)}

Out of 3 nights observed by us this blazar showed a hint of variablity on one 
night on applying the $F$-statistics, but remained non-variable on all the three nights,
according to the $C_{\rm eff}$ criterion (Table 3). However, on a longer term, we detected a clear 
fading by $\sim$0.1 mag between the first two epochs, 
which were separated by 10 days. This was followed by a phase of 
0.4 mag brightening over $\sim$ 2 months, when it was last monitored by us (Fig.\ 3).  }

\item{{J1221$+$2813 (W Comae/ON231, $z=0.102$)}

This is the nearest source in our sample and also the first intermediate
type BL Lac (IBL) to be detected at TeV energies. Gupta et al.\ (2008) 
have reported R-band monitoring of this blazar on 11 Jan.\ 2007 for 3.24
hours, but no INOV was detected. 
 We monitored this blazar on 4 epochs. It showed a confirmed
variablity on 3 epochs using $C_{\rm eff}$  and on all 4 epochs when the
$F$-test was applied (Table 3). In addition, a clear event of {\it internight} 
variation was seen when the source faded by $\sim$0.1mag between 19 and 20 Mar 2004. }

\item{{J1256$-$0547 (3C 279, $z=0.536$)}

This FSRQ of the OVV type was the first blazar to be detected as an
EGRET/$\gamma$-ray source (Hartman et al. 1992) and the first FSRQ to be detected 
at very high energies (VHE, i.e. $>$ 100 GeV) (MAGIC collaboration: Albert et al. 2008).
This source is also particularly interesting, because with a
redshift of 0.538, it is the most distant VHE blazar yet found
(Abdo et al.\ 2009); a strong absorption of its VHE emission due to extragalactic 
background light (EBL) is expected, yet not seen (e.g., Costamante et al.\ 2009).
Gupta et al. (2008) have reported $\it R$ band intra-night monitoring 
with the 1-m Yunnan observatory telescope on two nights in early 2007 
(for durations of 4 hour and 2.3 hours), but INOV was not detected.
A negative result was also reported by Romero et al.\ (2002) from
their V-band intranight monitoring lasting 3.8 hours on 06 Aug.\ 1999.

Application of either of the two statistics shows it to be a confirmed variable on 
all the 3 nights it was
monitored by us, attaining peak-to-peak INOV amplitudes of 4, 10 and 22 per cent,
respectively (Table 3). Further, the blazar brightened by $\sim$0.9 mag between the first two
epochs of monitoring that were separated by about 2 months, and was about 2.0 mag 
fainter when last observed a little more than 3 years later (Fig.\ 3).}

\item{{J1428$+$4240 (H1426$+$428, $z=0.129$)}

This weak TeV source has its synchrotron emission peak at the highest frequency
known for any  blazar; hence termed as ``extreme'' HBL 
(Aharonian et al.\ 2002; also, Costamante et al.\ 2001).
Its radio structure consists of a core surrounded by a faint halo
(Giroletti et al.\ 2004). The light curves presented
here are its first intranight optical monitoring observations. 
The $C_{\rm eff}$ values indicated it was non-variable on all  3 
nights we monitored it; however, on application of the $F$-test it was found to
be variable on 1 of the 3 nights (Table 3).  }

\item{{J1555+1111 (PG 1553$+$113, $z=0.360$)}

This most distant HBL with a firm TeV detection 
(Abdo et al. 2009) is known to show a highly time dependent 
variability behaviour at different frequencies.
Its radio structure is marked by an extremely large bend
(by $\sim 110^{\circ}$) of the parsec-scale jet towards the outer lobe
(Rector, Gabuzda \& Stocke 2003). It showed INOV on one of
the two nights it was monitored during mid-1999 as part of our INOV
programme (Stalin et al.\ 2005). Those data have been included
in our computation of the INOV duty cycle (Table 3). 
In our monitoring, confirmed variability was detected
on all 3 nights both using $C_{\rm eff}$ and $F$ statistics. 
In the longer term, a clear brightening by $\sim$ 0.25 mag was observed between the 
first two epochs of our monitoring, which were separated by almost a year.
A clear case of {\it internight} variability also occured when the source brightened by another 
$\sim$0.04 mag between 15 and 16 May 2010 (Fig.\ 3). }
\end{itemize}

\subsubsection{\bf Set 2} 

Set 2 consists of 1 HBL, 3 LBLs, 8 FSRQs and 1 BL Lac object of unspecified
HBL/LBL classification. Their INOV and LTOV data are available in the 
literature cited in Table 3. Only for OJ 287 do we present here new monitoring 
data taken by us on 2 nights (Table 3; Fig. 1).  

\begin{itemize}

\item{{J0238$+$1637 (AO 0235+164; $z =0.940$)}
  
This LBL had been monitored by Noble (1995) on 5 nights and later on 3 nights in 
the first part of our programme (Sagar et al.\ 2004). On all these 8 nights it 
was found to be a confirmed variable, with $\psi$ ranging between 5--20 per cent 
(Table 3). Romero et al.\ 
(2002) monitored it on another 6 nights and found $\psi$ to range between 
7--44 per cent. In addition, Goyal et al. (2011) have monitored this
blazar on one night and it showed a gradual decline by $\sim$ 7 per cent during 
the first 6 hours on that night. Thus, the INOV duty cycle of this BL Lac is 
essentially $100$ per cent! }

\item{{J0423$-$0120 (PKS 0420$-$01; $z = 0.915$)} 

This blazar has been newly monitored by us on 3
nights, covering a 7 years time span (Goyal et al.\ 2011)  It showed a confirmed variability on 
19 Nov.\ 2003, with an INOV amplitude of $\psi$ $\sim 2$ per cent and was a `probable 
variable' on 25 Oct.\ 2009 ($C$-test). The $F$-test shows it to 
be a confirmed variable on all the 3 nights. As for LTOV, it 
faded by $\sim$1.9-mag between the first 2 epochs of monitoring roughly a year
apart and then brightened by $\sim$0.8-mag when we last monitored it on 
25 Oct.\ 2009.} 

\item{{J0738$+$1742 (PKS 0735+178; $z > 0.424$)} 

This blazar is a rather unique case of relative intra-night quiescence that 
has persisted over the past two decades (Goyal et al.\ 2009; Britzen et al.\ 2010), although mild
INOV ($\psi$ $<$3 per cent) was detected on 5 out of the total 17 nights. 
On month-like timescale it has exhibited $\sim$0.5 mag variations (Sagar et al.\ 
2004; Ciprini et al.\ 2007; Goyal et al.\ 2009).  }

\item{{J0739$+$0137 (PKS 0736$+$01; $z = 0.191$)} 

Not only did this highly polarized CDQ remain a confirmed variable (by both 
$C$ \& $F$ statistics) on 2 of the 3 nights of our monitoring (Goyal et al.\ 2011), 
it also showed a clear {\it internight} variability, fading by $\sim$ 4 per cent 
between 5 and 6 Dec. 2005.}

\item{{J0854$+$2006 (OJ 287; $z = 0.306$)} 

This LBL was monitored by Sagar et al.\ (2004) on 4 nights and it 
showed confirmed INOV each time. It also showed large LTOV, first fading by 0.6 
mag over $\sim$ 1 year and then brightening by 1.63 mag on a 2-year time 
span. In Fig.\ 2 we present the light curves derived from our newly acquired 
data on 2 nights. The object showed a clear brightening by $7.5$ per cent in 4 hours
on 12 Apr 2005.
However, on 5 Feb 2005, it remained a non-variable (by the $C$-test), although 
a $F-$test showed it to be variable. }

\item{{J1058$+$0133 (PKS 1055$+$01; $z = 0.888$)} 

This FSRQ with a variable 
polarization was a confirmed variable on 2 of the 3 nights we monitored it 
(Goyal et al.\ 2011).  It also showed a gradual brightening by $\sim 0.2$ 
magnitude between 25 Mar.\ 2007 and 23 Apr.\ 2007.}

\item{{J1159$+$2914 (4C 29.45; $z = 0.729$)} 

Confirmed INOV was detected on all the 6 nights this FSRQ was monitored by 
Noble (1995), with $\psi$ ranging between 4 and 12 per cent (Table 3). 
It has also shown large variations on {\it internight} time scales. Firstly, 
it brightened by 6 per cent between 19 and 21 Jan.\ 1994 and then faded by 
$\sim$0.5 mag between 21 and 22 Jan.1994. It again brightened by 7 per cent
between 22 and 23 Jan.\ 1994 and was last found having faded by $\sim$0.3 mag on the
following night. }

\item{{J1217$+$3007 (B2 1215$+$30; $z = 0.130$)}

This blazar was monitored on 4 epochs by Sagar et al.\ (2004), showing
confirmed INOV on 2 epochs (Table 3). Variability was also detected on a
longer time scale. }

\item{{J1218$-$0119 (PKS 1216$-$01; $z = 0.554$)} 

INOV was detected on all the 4 nights this blazar was monitored by 
Sagar et al.\ (2004). On longer time scales, it showed a 2 per cent fading 
between the first 2 epochs of monitoring, followed by $\sim 11$ per cent fading 
over the next 2 days and, finally, 14 per cent brightening within the next 24 hours 
(Table 3).}

\item{{J1310$+$3220 (B2 1308$+$32; z = 0.997)} 

This FSRQ showed a confirmed INOV
($\psi$ up to 3.4 per cent) on one out of the 4 nights it was monitored by Sagar 
et al.\ (2004). In the longer term, it showed a strong variability (Stalin 
et al.\ 2004a). }

\item{{J1510$-$0906 (PKS 1510-08; $z = 0.360$)} 

This FSRQ showed confirmed INOV 
on 2 of the 3 nights it was monitored by us over the span of 5 years (Goyal et 
al., in prep.). In the long term, it brightened by $\sim 1.5$ mag over the 
course of 4 years and then faded by 0.25 mag over the next 19 days when we 
last monitored it on 20 May 2009. 
In earlier INOV campaigns by Romero et al.\ (2002) and Stalin et al.\ (2005),
this blazar was found to be non-variable.  }

\item{{J2225$-$0457 (3C 446; $z = 1.404$)} 

This blazar showed INOV on one of the 
2 nights it was monitored by Noble (1995) (Table 3). On 8 Oct 2010 when we 
monitored it (Goyal et al. 2011), it was found to be 
non-variable using the $C-$test, but variable by the $F$-test (Table 3). }

\item{{J2253$+$1608 (3C 454.3; $z = 0.859$)} 

INOV was detected on both the nights it was monitored by Noble (1995) (Table 3).}

\end{itemize}

\section{Discussion}

The fairly large size of the TeV blazar sample covered in the present study
and the fact that all the key blazar subclasses, namely HBL, IBL, LBL and FSRQ
(Sect.\ 2; Table 1) are included in the sample, reassures us that the INOV 
results reported here should be representative of TeV blazars. An explicit
goal of the present work, the first systematic study to define the INOV 
characteristics of TeV blazars, was to search for variations on sub-hour 
time scales. As pointed out in Sect. 1, TeV blazars are 
particularly promising targets to look for such ultra-rapid optical 
continuum variability, since their
innermost jets are thought to have extremely large bulk Lorentz factors, 
$\Gamma \ga 50 - 100$. Despite examining the high-quality intranight 
monitoring observations of all 22 TeV blazars in our sample, taken on 116 nights over a total 
duration of 677 hours, no clear event on sub-hour time scales was found, even 
though the data sampling rate (typically, once every 10 minutes or so) was 
adequately dense to have revealed any such cases. This demonstrates 
that the occurrence of
optical continuum variability on sub-hour like time scales must be exceedingly
rare, at least for amplitudes above $\sim$ 1 per cent, the detection 
threshold typically reached in the deepest INOV searches made so far. 
It may nonetheless be noted that our intranight DLCs (Fig.\ 1) do exhibit a few clear 
cases of rather sharp ``bumps'', notably in the cases of J0721$+$7120 
(1 Feb.\ 2005), J1221$+$2813 (20 Mar.\ 2004) and J1256$-$0547 (26 and 28 Jan.\ 
2006).

As mentioned in Sect.\ 1, parsec scale radio jets in $\gamma-$ray (EGRET) 
blazars are known to show faster apparent (superluminal) motion as well as 
relatively higher brightness temperatures in their cores, compared to 
non-EGRET blazars (e.g., Kellermann et al.\ 2004; Jorstad et al. 2001; 
Taylor et al.\ 2007; Lister et al.\ 2009a). 
Indeed, in an earlier part of our INOV programme, evidence was reported
for EGRET detected blazars to exhibit a stronger INOV, again suggesting 
a link between INOV and the jet speed (Stalin et al.\ 2005). An independent
evidence for such a physical link has emerged in the present study, from a comparison of the INOV 
duty cycles determined for the two major subclasses of BL Lacs, namely `low-peaked-BL 
Lacs (LBL)' and `high-peaked BL Lacs (HBLs)'. The 
synchrotron emission from LBLs peaks in the near-infrared/optical domain 
while for HBLs the synchrotron peak falls in the UV/soft X-ray regime (Urry \& Padovani 
1995 and references therein). Of the two classes, LBLs are known to display 
greater activity and this is generally attributed to their emission being 
dominated by faster jets which are probably also better aligned to our
direction (e.g., Ghisellini \& Maraschi 1989; Sambruna, Maraschi \& Urry 1996; 
 Kharb, Gabuzda \& Shastri 2008). Consistent with this picture, it has been 
found that LBLs display stronger INOV than do HBLs; the duty cycle has been 
estimated to be $\sim$70 per cent for LBLs and $\sim$30--50 per cent for HBLs 
(e.g., Romero et al., 1999, 2002; see also, Heidt \& Wagner 1998).

The present study allows us to check, for the first time, whether the 
difference persists even when the two blazar subclasses are subjected to the 
condition of TeV detection. From Table 1, our sample consists of 7 HBLs
(with the lower hump of the SED peaking above $10^{15}$ Hz) and 15 other blazars which can be 
together termed 
LBLs  (see, Abdo et al.\ 2010b; Li et al.\ 2010) including J1218$-$0119 which is a 
flat-spectrum core dominated quasar.
The estimated INOV duty cycles are 38 per cent for HBLs (26 nights' 
data) and 63 per cent for LBLs (90 nights' data)(Table 3). If only the cases of
INOV amplitudes $\psi > 3$ per cent  are considered, the duty cycle is 22 per cent for HBLs and 
50 per cent for LBLs. Both the estimates for the TeV detected
LBLs are in close agreement with those reported in Gopal-Krishna et al. (2003), 
for radio selected BL Lacs (believed to be predominently LBLs). 
Figure 4 shows the histograms of INOV DCs for HBLs and LBLs in our sample.
The K-S test rejects at the $99$ per cent confidence level the hypothesis 
that the two 
distributions are drawn from the same parent population. Thus, it is evident 
that the strong tendency for HBLs to show a milder INOV, vis-\'{a}-vis LBLs, 
persists even if the comaprison is restricted to their TeV detected subsets. This 
result  must be accounted for in physical models of INOV and TeV emission from blazars.

A striking outcome of our search for ultra-rapid INOV is 
the detection of a curious, sharp feature 
in the light curve of the HBL J1555$+$1111 (on 24 June 2009; Fig.\ 1), the 
most optically luminous member of Set 1 (Table 1).  Intriguingly, instead of being a 
flare, as is more typical, this essentially symmetric feature is a clear ``dip'', with an amplitude of 
$\sim$ 4.2 per cent. Its initial, falling side extends 0.5 hour 
and the rising side is of a slightly longer duration. We treat this detection 
as robust, since (i) The profile of the ``dip'' is well resolved (8 data 
points); (ii) its peak amplitude is $>$ 20 times the rms noise of individual 
data points; (iii) the variation is visible with equal clarity and amplitude on the DLCs 
of the blazar against all the 3 comparison stars; (iv) the 3 comparison stars
are all within 0.8 magnitude of the blazar; (v) all 3 comparison stars
remained rock-steady through the monitoring duration; and (vi) the atmospheric
seeing too remained steady throughout the monitoring session (Fig.\ 1).
All these points, along with the fact that in the DLCs, the brightness levels
seen immediately prior to and just after the dip are fairly steady and well
matched, {\it make this feature by far the most credible, if not the only,
case of an emission dip (on an hour-like time scale) detected in the
optical continuum light received from a blazar}. In addition, the  blazar was 
substantially brighter on both other nights it was monitored by us (Table 3), 
making it quite unlikely that the elevated flux levels seen just before and 
after the dip in the light curves of 24 June 2009 are a result of 
multiple flares occuring at those times. 

The robust detection of the dip, combined with its temporal sharpness, calls 
for an explanation. A plausible scenario for this rare feature would be that 
the optical continuum from the jet was temporarily absorbed by a foreground
dusty cloud in the nuclear region. An emerging picture of AGN 
broad emission line clouds suggests they form at the distances from the central continuum 
source where the temperature becomes low enough for dust formation (i.e., 
$\sim 10^3$ K) and these dusty clouds then form a dust-driven wind (e.g., Czerny 
\& Hryniewicz 2010; also, Krongold, Binnete \& Hernandez-Ibarra 2010). 
Recalling that the low-level detection (or non-detection) of emission lines 
in the case of BL Lacs could well be due to a paucity of energetic photons 
to ionise the gas clouds existing around the central engine (e.g., Ghisellini et al. 
2009), it is plausible that a dusty cloud, or a stream of such 
clouds happens to be on our line of sight to a superluminally moving optical 
synchrotron `knot' in the jet.  To see if this
could explain the ``dip'', we need an estimate of the size of 
an optically emitting knot in blazar jets.  Here a reasonable expectation would be 
that optical knots already exist in the jet before the point where the knot
becomes visible in the radio band (e.g.\ Marscher et al.\ 2008); 
that point typically occurs at a distance, $l_{rad}$ $\approx 10^3$ times 
the gravitational radius of the central supermassive black hole (Lobanov 
2010). For a typical supermassive black hole of mass $10^8 M_{\odot}$, the 
radio visibility point would thus be about $10^{16.5}$ cm away from the nucleus, 
from which a size of around $10^{15}$ cm can be reasonably inferred for the 
optical knot in the jet. The observed time scale of the optical continuum dip
($\sim 10^3$ sec), if interpreted in terms of an occultation of the superluminal knot by a dusty 
gaseous cloud (see above), would then require a relative transverse speed of 
order of $30c$ (see Gopal-Krishna \& Subramanian 1991 for such a scenario). The 
corresponding bulk Lorentz factor of the inner jet, $\Gamma \geq 30$ 
for viewing angle $\la$ 2$^\circ$, 
is not unreasonable for a TeV blazar (Sect. 1). One prediction of this scenario 
is that any such rapid intensity dips should appear more pronounced in B-band 
light curves, because of their greater sensitivity to dust obscuration, as compared 
to the R-band light curves.

Finally, searches for minute-like time scales in the optical light curves of
blazars, a key motivation for the present work, have acquired added
significance in the present Fermi/LAT era (Atwood et al.\ 2009). A few
recent studies of blazars have revealed a tight physical relationship
between $\gamma-$ray and optical flaring events, strongly suggesting
a spatial coincidence between their origins, which could well be in the
jet's knot crossing a standing shock, located up to several parsecs
from the central engine (see, e.g., Agudo et al.\ 2010, and references
therein).
More specifically, a recent study of blazars
(Ammando et al.\ 2010) has revealed that polarized optical emission from
their jets is spatially coincident with the site of the TeV flares
observed on minute-like time scales. This emerging scenario provides further motivation
for more concerted efforts to search for optical variability of TeV blazars 
on minute-like time scales.

\section*{Acknowledgments}
The authors are thankful to the anonymous referee for helpful
suggestions. This research has made use of NASA/IPAC Extragalactic Database (NED),
which is operated by the Jet Propulsion Laboratory, California Institute
of Technology, under contract with National Aeronautics and Space
Administration. We thank N. K. Chakradhari for carrying out observations for us. 
The authors wish to acknowledge the support received from the staff of the
IAO and CREST of IIA and the IUCAA-Girawali Observatory (IGO) of IUCAA.


\clearpage
\newpage
\begin{landscape}
\begin{table}
\centering
\caption{The sample of 22 TeV blazars studied in the present work}

\begin{tabular}{ccccccccccccccc}\\
\hline
IAU name & Other name   & Type & R.A.(J2000) & Dec(J2000)&{\it B}& $M_{B}$& z &$P_{op}(\%)$&$\alpha_{r}$&$P_{c}^{5 GHz}$ & $P_{ext}^{5 GHz}$ &$log f_c$&  $log R^*$&  $\beta_{app}$ \\
         &       &       & (h m s)    &($ ^{\circ}$ $ ^{\prime}$ $ ^{\prime\prime }$) &        &   &       &     &        & (W/Hz)& (W/Hz) &        &     & (max)   \\
  (1)    &  (2)  & (3)   & (4)        & (5)   &  (6)   &(7)& (8)        &    (9) &   (10)   &(11)  & (12) & (13)   &(14) &(15)\\
\hline
\multicolumn{15}{|l|}{Set 1}\\ \hline
J0222$+$4302&3C 66A    & IBL  & 02 22 39.6  &$+$43 02 08  & 15.71  & $-$24.96  & 0.444       &16.8$^a$  & $-$0.20$^\S$  &$6.5\times10^{26}$$^h$   & $1.1\times10^{27}   $&-0.22 & 2.98  & 5.8$^n$  \\
J0721$+$7120&S5 0716+71  & IBL  & 07 21 53.3  &$+$71 20 36  & 15.50  & $-$24.73  & 0.310       &29.0$^b$  & $+$0.36   &$1.0\times10^{27}$$^i$   & $4.5\times10^{26}   $& 0.36 & 2.43  & 10.07$^o$ \\
J0809$+$5218&1ES 0806+524 & HBL & 08 09 49.2  &$+$52 18 58  & 15.30  & $-$23.25  & 0.138       &          & $-$0.25  &$5.6\times10^{24}$$^j$   & $2.7\times10^{24}   $& 0.30 & 1.79  &          \\
J1015$+$4926&GB 1011+496 & HBL  & 10 15 04.2  &$+$49 26 01  & 16.56  & $-$22.88  & 0.200       &          & $-$0.26  &$2.0\times10^{25}$$^k$   & $1.8\times10^{25}   $& 0.03 & 2.59  &          \\
J1221$+$3010&PG 1218+304 & HBL  & 12 21 21.9  &$+$30 10 37  & 16.50  & $-$22.72  & 0.182       &6.6$^e$   & $-$0.08  &$5.4\times10^{24}$$^j$   & $1.3\times10^{24}   $& 0.58 & 1.85  &        \\
J1221$+$2813&ON 231 & HBL  & 12 21 31.7  &$+$28 13 58  & 16.81  & $-$21.21  & 0.102       &4.3$^e$   & $-$0.09$^\S$  &$1.5\times10^{25}$$^i$   & $3.1\times10^{24}   $& 0.70 & 3.01  & 3.2$^q$ \\
J1256$-$0547&3C 279    & FSRQ & 12 56 11.1  &$-$05 47 21  & 18.01  & $-$23.24  & 0.538       &44.0$^b$  & $+$0.48$^\S$  &$3.2\times10^{28}$$^i$   & $1.0\times10^{28}   $& 0.50 & 4.51  & 20.58$^o$\\
J1428$+$4240&H 1426+428 & HBL  & 14 28 32.7  &$+$42 40 21  & 16.95  & $-$21.60  & 0.129       &2.5$^d$   & $-$0.58   &$1.5\times10^{24}$$^j$   & $7.2\times10^{23}   $& 0.34 & 1.96  & 2.09$^r$ \\
J1555$+$1111&PG 1553+11 & HBL  & 15 55 43.1  &$+$11 11 24  & 15.00  & $-$25.42  & 0.360$^\ddag$&3.5$^g$  & $+$0.54  &$1.5\times10^{26}$$^m$   & $1.9\times10^{25}   $& 0.88 & 1.85  &         \\\hline
\multicolumn{15}{|l|}{Set 2}\\ \hline
J0238$+$1637&AO 0235+164 & LBL  & 02 38 38.9  &$+$16 37 00  & 16.46  & $-$25.47  & 0.940       &43.9$^b$  & $+$0.71$^\S$  &$1.3\times10^{28}$$^i$   & $2.1\times10^{27}   $& 0.80 & 3.12  &         \\
J0423$-$0120&PKS 0420-01  & FSRQ & 04 23 15.8  &$-$01 20 33  & 17.50  & $-$24.17  & 0.915       &20.0$^b$  & $+$0.19$^\S$  &$8.2\times10^{28}$$^i$   & $4.4\times10^{28}   $& 0.27 & 3.69  & 7.35$^o$ \\
J0738$+$1742&PKS 0735+17  & LBL  & 07 38 07.4  &$+$17 42 19  & 16.76  & $-$24.04  &$>$0.424     &36.0$^b$  & $-$0.28$^\S$  &$1.5\times10^{27}$$^i$   & $3.6\times10^{26}   $& 0.61 & 3.40  & 11.8$^p$ \\
J0739$+$0137&PKS 0736+01  & FSRQ & 07 39 18.0  &$+$01 37 04  & 16.90  & $-$21.96  & 0.191       &5.6$^c$   & $-$0.10$^\S$  &$2.2\times10^{26}$$^i$   & $1.3\times10^{25}   $& 1.20 & 3.45  & 14.44$^o$ \\
J0854$+$2006&OJ 287   & LBL  & 08 54 48.8  &$+$20 06 30  & 15.91  & $-$24.30  & 0.306       &37.2$^b$  & $+$0.20$^\S$  &$1.6\times10^{27}$$^i$   & $6.2\times10^{26}   $& 0.41 & 2.91  & 11.70$^o$ \\
J1058$+$0133&PKS 1055+01  & FSRQ & 10 58 29.6  &$+$01 33 58  & 18.74  & $-$23.34  & 0.888       &5.0$^d$   & $+$0.06$^\S$  &$3.9\times10^{28}$$^i$   & $1.1\times10^{28}   $& 0.54 & 4.26  & 11.0$^o$\\
J1159$+$2914&4C 29.45 & FSRQ & 11 59 31.9  &$+$29 14 45  & 14.80  & $-$27.00  & 0.729       &28.0$^b$   & $-$0.34$^\S$  &$1.3\times10^{28}$$^i$   & $3.6\times10^{27}   $& 0.57 & 2.60  & 24.8$^o$\\
J1217$+$3007&B2 1215+30  & HBL  & 12 17 52.0  &$+$30 07 01  & 16.07  & $-$22.45  & 0.130       &8.0$^e$   & $-$0.12$^\S$  &$1.3\times10^{25}$$^j$   & $8.8\times10^{24}   $& 0.17 & 2.60  &        \\
J1218$-$0119&PKS 1216-010 & BL$^\pounds$     & 12 18 34.9  &$-$01 19 54.0  & 16.17  & $-$25.14  & 0.554       &6.9$^f$   & $+$0.01$^\S$  &$4.3\times10^{27}$$^l$   & $2.5\times10^{27}   $& 0.22 & 3.01  &         \\
J1310$+$3220&B2 1308+32  & FSRQ & 13 10 28.7  &$+$32 20 44  & 15.61  & $-$26.69  & 0.997       &28.0$^b$  & $-$0.09$^\S$ &$3.0\times10^{28}$$^i$   & $9.9\times10^{27}   $& 0.49 & 2.71  & 27.2$^o$ \\
J1512$-$0906&PKS 1510-08&FSRQ & 15 12 50.5  &$-$09 06 00  & 16.74  & $-$23.49  & 0.360       &7.8$^b$   & $-$0.10$^\S$  &$1.7\times10^{27}$$^i$   & $7.7\times10^{25}   $& 1.35 & 3.48  & 20.2$^o$ \\
J2225$-$0457&3C 446    & FSRQ & 22 25 47.2  &$-$04 57 01  & 18.83  & $-$23.66  & 1.404        &17.3$^b$ & $-$0.13$^\S$ &$1.3\times10^{29}$$^i$   & $6.2\times10^{28}   $& 0.34 & 4.58  & 17.34$^o$\\
J2253$+$1608&3C 454.3  & FSRQ & 22 53 57.7  &$+$16 08 53  & 16.57  & $-$25.11  & 0.859        &16.0$^b$ & $-$0.28$^\S$ &$6.7\times10^{28}$$^i$   & $1.2\times10^{28}   $& 0.74 & 3.99  & 14.19$^o$\\

\hline
\end{tabular}
Columns: (1) source name; (2) most popular name as given in
V\'eron \& V\'eron (2006); (3) classification into low-, intermediate-, or high-frequency peaked blazars
or flat spectrum radio quasar (FSRQ); (4) right ascension;
(5) declination;
(6) apparent B-magnitude; (7) absolute B-magnitude; (8) redshift;
(9) the measured optical polarization; (10) radio spectral index;
(11) radio core luminosity; (12) extended emission radio luminosity; 
(13) radio core dominance parameter $f_c$ (see text);
(14) radio loudness $R^*$; 
(15) the fastest apparent speed observed in the parsec-scale jet (in units of the speed of light).\\
Footnotes : \\
Column 3 : Abdo et al.\ (2010a), $^\pounds$ SED classification is not available. \\
Column 8 : $^\ddag$ reference for redshift: Rector et al.\ (2003), see also, Treves, Falomo \& Uslenghi (2007)\\
Column 9 : (a) Takalo, Silanp{\"a}{\"a} \& Nilsson (1994); (b) Fan et al.\ (1997); (c) Stockman, Moore \& Angel (1984); 
(d) Jannuzi, Smith \& Elston (1993) ; 
(e) Wills et al.\ (1992); (f) Visvanathan \& Wills (1998); (g) Andruchow, Romero \& Cellone (2005). \\
Column 10 : {$^\S$} is the radio spectral index derived using simultaneous flux measurements from Kovalev et al.\ (1999)
by means of least square method ($S_{\nu} \propto \nu^{\alpha}$). The remaining values of $\alpha_r$ have been calculated
using 6 and 20 cm. flux densities from V\'eron \& V\'eron (2006). \\
Column 11 : Reference for VLBI flux: (h) Marscher et al.\ (2002); (i) Kovalev et al.\ (2005); (j) Giroletti et al.\ (2006); (k) Helmboldt et al.\ (2007); 
                          (l) Wehrle, Morabito \& Preston (1984); (m) Rector et al.\ (2003). \\ 
Column 14 : $R^*$ is the {\it K}-corrected rest frame ratio of the 5 GHz to 2500
\AA~flux densities, following Stocke et al. (1992); reference for the radio flux is V\'eron \& V\'eron (2006).\\
Column 15 : Reference for $\beta_{app}$: (n) Britzen et al.\ (2008); (o) Lister et al.\ (2009b); (p) Britzen et al.\ (2010); (q) Kellermann et al.\ (2004); (r) Piner et al.\ (2008).\\

\end{table}
\end{landscape}

\clearpage
\newpage
\begin{table} 
\caption{Positions and apparent magnitudes$^*$ of the TeV blazars and the comparison stars.} 
\begin{tabular}{lcccccc}\\
\hline
Source &  Set No.  & R.A.(J2000) &Dec(J2000) &  {\it B}$^*$ & {\it R}$^*$  & {\it B-R}  \\
      &     & (h m s)     & ({$^\circ$} {$^\prime$} {$^{\prime\prime}$})  & (mag)     & (mag)         & (mag)    \\\hline 
J0222$+$4302&  1  &02 22 39.61 & $+$43 02 07.9 & 14.94 & 14.35&  0.59    \\
S1          &     &02 22 28.41 & $+$43 03 40.9 & 14.59 & 14.00&  0.59    \\
S2          &     &02 22 20.45 & $+$42 57 18.7 & 14.74 & 13.66&  1.08    \\
S3          &     &02 22 39.09 & $+$42 57 05.5 & 14.69 & 13.94&  0.75    \\
J0238$+$1637&  2  &02 38 38.92 & $+$16 36 59.2 & 18.65 & 15.92 &2.73\\
         S1 &     &02 38 56.00 & $+$16 37 43.0 & 17.43 & 16.60 &0.83\\
         S2 &     &02 38 38.52 & $+$16 40 05.3 & 18.37 & 16.61 &1.76\\
         S3 &     &02 38 22.25 & $+$16 39 41.8 & 17.37 & 16.22 &1.15\\
J0721$+$7120&  1  &07 21 53.39 & $+$71 20 36.7 & 15.15 & 14.27&  0.88    \\
S1          &     &07 22 12.58 & $+$71 21 15.2 & 14.49 & 13.78&  0.71    \\
S2          &     &07 21 54.31 & $+$71 19 21.2 & 14.46 & 13.67&  0.79    \\
S3          &     &07 21 13.95 & $+$71 17 10.0 & 14.45 & 13.55&  0.90    \\
J0809$+$5218&  1  &08 09 49.20 & $+$52 18 58.4 &       &      &          \\
S1          &     &08 09 43.90 & $+$52 18 09.5 & 16.17 & 15.47&  0.70    \\
S2          &     &08 10 16.45 & $+$52 19 16.9 & 17.85 & 16.10&  1.75    \\
S3          &     &08 09 52.68 & $+$52 16 15.1 & 15.25 & 14.73&  0.52    \\
J1015$+$4926&  1  &10 15 04.13 & $+$49 26 00.8 & 15.32 & 14.58&  0.74    \\
          S1&     &10 15 29.71 & $+$49 30 41.5 & 15.27 & 14.84& 0.43     \\
          S2&     &10 15 37.87 & $+$49 28 19.2 & 14.92 & 14.56& 0.36     \\
          S3&     &10 15 08.03 & $+$49 25 42.3 & 14.27 & 13.55& 0.72     \\
          S4&     &10 15 39.84 & $+$49 29 25.7 & 16.70 & 14.95& 1.75     \\
J1221$+$3010&  1  &12 21 21.93 & $+$30 10 37.2 & 16.13 & 14.93& 1.20     \\
          S1&     &12 21 22.62 & $+$30 09 53.8 & 16.72 & 15.43& 1.29     \\
          S2&     &12 21 23.08 & $+$30 10 38.9 & 17.04 & 15.42& 1.62     \\
          S3&     &12 21 37.05 & $+$30 10 18.3 & 16.28 & 15.52&  0.76    \\
J1221$+$2813&  1  &12 21 31.69 & $+$28 13 58.4 & 14.65 & 14.24&  0.41    \\
S1          &     &12 21 26.01 & $+$28 12 30.9 & 16.94 & 15.91&  1.03    \\
S2          &     &12 21 13.86 & $+$28 13 04.5 & 13.61 & 12.88&  0.73    \\
S3          &     &12 21 11.81 & $+$28 11 53.7 & 16.13 & 15.00&  1.13    \\
J1256$-$0547&  1  &12 56 11.19 & $-$05 47 21.5 & 17.39 & 15.87&  1.52    \\
         S1 &     &12 56 26.61 & $-$05 45 22.8 & 15.22 & 14.75&  0.47    \\
         S2 &     &12 55 58.00 & $-$05 44 18.9 & 16.19 & 15.30&  0.89    \\
         S3 &     &12 56 14.48 & $-$05 46 47.8 & 16.39 & 15.43&  0.96    \\
J1428$+$4240&  1  &14 28 32.62 & $+$42 40 21.4 &       &      &          \\
S1          &     &14 28 08.06 & $+$42 40 37.4 & 16.62 & 16.23&  0.39    \\
S2          &     &14 28 07.38 & $+$42 44 20.0 & 16.16 & 15.74&  0.42    \\
S3          &     &14 28 16.05 & $+$42 40 39.9 & 16.77 & 16.56&  0.21    \\
J1555$+$1111&  1  &15 55 43.05 & $+$11 11 24.3 & 14.30 & 13.99&  0.31    \\
S1          &     &15 55 46.08 & $+$11 11 19.6 & 14.52 & 13.62&  0.90    \\
S2          &     &15 55 52.17 & $+$11 13 18.5 & 14.47 & 13.56&  0.91    \\
S3          &     &15 55 40.77 & $+$11 04 44.7 & 14.46 & 13.56&  0.90    \\
S4          &     &15 56 06.02 & $+$11 13 44.9 & 15.44 & 14.54&  0.90    \\
J0854$+$2006&  2  &08 54 48.68 & $+$20 06 30.8 & 15.95 & 15.56&  0.40    \\
S1          &     &08 54 53.36 & $+$20 04 45.1 & 15.25 & 13.97&  1.28    \\
S2          &     &08 54 41.29 & $+$20 06 43.2 & 16.86 & 15.60&  1.26    \\
S3          &     &08 54 55.19 & $+$20 05 42.4 & 15.83 & 14.94&  0.89    \\
\hline
\hline
\end{tabular}

$^*$Monet et al.\ (2003)
\end{table}

\newpage

\begin{deluxetable}{cccccccccccccc}
\tablecolumns{14}
\setlength{\tabcolsep}{0.03in}
\tablewidth{0pc}
\tabletypesize{\tiny} 
\tablecaption{Summary of observations and INOV results}
\tablehead{
Source &  Set No.  & Epoch &Tel.      &Fil. &Dur.&$N_{points}$&$\sigma $&$\psi$ &   $ C_{\rm eff}$  & Status$^\dag$& F-value(DoF$^\ddag$)     &  Status$^\dag$   & Ref.$^\pounds$  \\
       &        &  dd.mm.yy& used     &       & (hours)&           & (\%)    &  (\%)  &               &   ($C$-test)       &     & ($F$-test)       &    \\
        (1) & (2) & (3) & (4) & (5) & (6) & (7) & (8) & (9) & (10) & (11) & (12) & (13) & (14) \\ }
\startdata

J0222$+$4302 &  1 &14.11.98 & ST  & R &  6.5  & 118  &        & 5.4   &   6.0    & V    &           &      & (a)\\
             &    &13.11.99 & ST  & R &  5.7  & 123  &        & 5.5   &$>$6.6    & V    &           &      & (a)\\
             &    &24.10.00 & ST  & R &  9.1  &  73  &        & 4.3   &   5.8    & V    &           &      & (a)\\
             &    &26.10.00 & ST  & R & 10.1  &  82  &        & 3.2   &   3.5    & V    &           &      & (a)\\
             &    &01.11.00 & ST  & R &  9.0  & 103  &        & 2.2   &   2.9    & V    &           &      & (a)\\
             &    &24.11.00 & ST  & R &  5.1  &  71  &        &       &          &N     &           &      & (a)\\
             &    &01.12.00 & ST  & R &  5.1  &  59  &        & 8.0   &$>$6.6    & V    &           &      & (a)\\
             &    &27.09.09 & ST  & R &  4.1  &  34  &  0.28  & 1.0   & 1.6      & N    & 1.15(33)  & N    & (b)\\
             &    &28.09.09 & ST  & R &  6.2  &  38  &  0.19  & 1.1   &  2.6     &  V   & 3.75(37)  & V    & (b)\\
             &    &         &     &   &       &      &        &       &          &      &           &      &    \\
J0238$+$1637 &  2 &27.10.90 &     & V &4.5    &      &  1.00  &  5.0  &          & V    &           &      &(c) \\
             &    &29.10.90 &     & V &8.3    &      &  1.20  &  18.0 &          & V    &           &      &(c) \\
             &    &05.11.91 &     & R &6.3    &      &  1.20  &  20.0 &          & V    &           &      &(c) \\
             &    &07.11.91 &     & R &8.0    &      &  1.20  &   5.0 &          & V    &           &      &(c) \\
             &    &08.11.91 &     & R &9.4    &      &  1.30  &   16.0&          & V    &           &      &(c) \\
             &    &03.11.99 &     & V &6.7    &      &  1.40  &   27.3& 10.0     & V    &           &      &(d) \\
             &    &04.11.99 &     & V &6.6    &      &  1.20  &   24.5&  6.1     & V    &           &      &(d) \\
             &    &05.11.99 &     & V &7.0    &      &  1.20  &   34.5&  8.9     & V    &           &      &(d) \\
             &    &06.11.99 &     & V &6.7    &      &  0.70  &   11.0&  4.4     & V    &           &      &(d) \\
             &    &07.11.99 &     & V &6.6    &      &  0.90  &   44.3&  14.4    & V    &           &      &(d) \\
             &    &12.11.99 &  ST & R &6.6    & 39   &        &   12.8& 6.6      & V    &           &      &(a) \\
             &    &14.11.99 &  ST & R &6.2    & 34   &        &   10.2& 3.2      & V    &           &      &(a) \\
             &    &22.10.00 &  ST & R &7.9    & 39   &        &    7.6& 2.6      & V    &           &      &(a) \\
             &    &22.12.00 &     & V &7.2    &      &  0.70  &    7.0&   3.3    & V    &           &      &(d) \\
             &    &18.11.03 & HCT & R &7.4    & 39   &  0.40  &    6.8& $>$5.54  & V    & 58.71(38) & V    & (e) \\
             &    &         &     &   &       &      &        &       &          &      &           &      &   \\
J0423$-$0120 &  2 &19.11.03 &  ST & R &6.3    & 36   &  0.25  &  1.6  & 3.60     & V    & 7.81(35)  &  V   &(e) \\
             &    &08.12.04 &  ST & R &6.0    & 11   &  0.26  &  1.8  & 0.96     & N    & 10.51(10) &  V   &(e) \\
             &    &25.10.09 &  ST & R &4.0    & 18   &  0.38  &  2.9  & 2.00     & PV   & 6.45(17)  &  V   &(e) \\
             &    &         &     &   &       &      &        &       &          &      &           &      &    \\
J0721$+$7120 &  1 &23.03.04 &     & R &  7.0  &      &        & 6.0   &          & V    &           &      & (f)\\
             &    &01.02.05 & ST  & R &  1.5  &  24  &  0.25  & 3.1   &  5.1     & V    & 22.15(23) & V    & (b)\\
             &    &12.01.07 &     & R &  3.6  & 185  &  0.70  &  6.3  &  2.8     & V    &           &      & (g)\\
             &    &20.03.07 &     & R &  4.2  & 195  &  0.50  &       &  1.6     & N    &           &      & (g)\\
             &    &         &     &   &       &      &        &       &          &      &           &      &    \\
J0738$+$1742 &  2 &26.12.98 &  ST & R &7.8    & 49   &        &1.8    &   1.13   &   N  &           &      & (a)\&(h) \\
             &    &30.12.99 &  ST & R &7.4    & 64   &        &1.0    &   0.61   &   N  &           &      & (a)\&(h) \\
             &    &25.12.00 &  ST & R &6.0    & 42   &        &1.6    &   1.02   &   N  &           &      & (a)\&(h) \\
             &    &25.12.01 &  ST & R &7.3    & 43   &        &1.0    &   2.8    &   V  &           &      & (a)\&(h) \\
             &    &20.12.03 & HCT & R &5.8    & 36   &  0.23  &1.0    &   1.78   &   N  & 1.86(35)  &   N  & (h)    \\
             &    &10.12.04 &  ST & R &5.8    & 28   &  0.16  &1.3    &   3.00   &   V  & 8.40(27)  &   V  & (h)    \\
             &    &23.12.04 &  ST & R &5.0    & 11   &  0.10  &1.2    &   3.10   &   V  & 17.83(10) &   V  & (h)    \\
             &    &02.01.05 &  ST & R &4.9    & 20   &  0.22  &0.8    &   0.97   &   N  & 1.42(19)  &   N  & (h)    \\
             &    &05.01.05 &  ST & R &5.8    & 23   &  0.15  &1.0    &   2.25   &  PV  &  7.00(22) &   V  & (h)    \\
             &    &09.01.05 &  ST & R &6.7    & 28   &  0.19  &1.3    &   3.20   &   V  &  9.78(27) &   V  & (h)    \\
             &    &09.11.05 &  ST & R &3.8    & 17   &  0.17  &0.7    &   2.00   &  PV  &  3.35(16) &   PV & (h)    \\
             &    &16.11.06 &  ST & R &4.5    & 19   &  0.29  &1.1    &   0.95   &   N  &  4.35(18) &  V   & (h)    \\
             &    &29.11.06 &  ST & R &5.8    & 26   &  0.19  &1.0    &   0.83   &   N  &  3.42(25) &  V   & (h)    \\
             &    &17.12.06 &  ST & R &5.6    & 24   &  0.19  &0.9    &   1.06   &   N  &  5.72(23) &  V   & (h)    \\
             &    &15.12.07 &  ST & R &6.6    & 28   &  0.29  &1.9    &   3.53   &   V  &  10.37(27)&  V   & (h)    \\
             &    &16.12.07 &  ST & R &6.6    & 27   &  0.19  &1.0    &   1.45   &   N  &  3.87(26) &  V   & (h)    \\
             &    &22.11.08 &  ST & R &5.6    & 27   &  0.19  &0.8    &   0.33   &   N  &  1.84(26) &  N   & (h)    \\
             &    &         &     &   &       &      &        &       &          &      &           &      &    \\
J0739$+$0137 &  2 &05.12.05 & HCT & R &5.3    & 10   &  0.38  &  3.4  & 2.93     & V    &  36.84(9) &  V   &(e) \\
             &    &06.12.05 & HCT & R &6.0    & 09   &  0.54  &  3.1  & 3.50     & V    &  6.23(8)  &  V   &(e) \\
             &    &09.12.05 & HCT & R &5.5    & 14   &  0.54  &  1.3  & 0.38     & N    &  1.45(13) &  N   &(e) \\
             &    &         &     &   &       &      &        &       &          &      &           &      &    \\
J0809$+$5214 &  1 &04.02.05 & HCT & R &  6.8  & 27   & 0.20   & 1.2   &  2.31    & PV   & 1.86(26)  & N    &(b) \\
             &    &05.12.05 & HCT & R &  4.3  & 08   & 0.24   & 0.8   &  0.59    & N    & 3.40(7)   & N    &(b) \\
             &    &08.12.05 & HCT & R &  4.9  & 14   & 0.11   & 0.4   &  0.15    & N    & 1.45(13)  & N    &(b) \\
             &    &09.12.05 & HCT & R &  4.8  & 12   & 0.24   & 0.5   &  0.32    & N    & 1.27(11)  & N    &(b) \\
             &    &         &     &   &       &      &        &       &          &      &           &      &    \\
J0854$+$2006 &  2 &29.12.98 &  ST & R &  6.8  & 19   &        & 2.3   &  2.80    &  V   &           &      & (a)   \\
             &    &31.12.99 &  ST & R &  5.6  & 29   &        & 3.8   &  6.50    &  V   &           &      & (a)   \\
             &    &28.03.00 &  ST & R &  4.2  & 22   &        & 5.0   &  5.80    &  V   &           &      & (a)   \\
             &    &17.02.01 &  ST & R &  6.9  & 48   &        & 2.8   &  2.70    &  V   &           &      & (a)   \\
             &    &05.02.05 & HCT & R &  7.7  & 40   &  0.23  & 0.8   &  1.69    &  N   & 2.57(39)  & V    & (b)   \\
             &    &12.04.05 &  ST & R &  4.1  & 54   &  0.32  & 7.5   &  5.20    &  V   & 71.39(54) & V    & (b)   \\
             &    &         &     &   &       &      &        &       &          &      &           &      &    \\
J1015$+$4926 &  1 &06.02.10 &  ST & R &  5.5  & 24   &  0.25  & 1.0   &     0.98 & N    & 2.08(23)  & N    &(b) \\
             &    &19.02.10 &  ST & R &  5.6  & 41   &  0.30  & 1.8   &     2.17 & PV   & 5.06(40)  & V    &(b) \\
             &    &07.03.10 &  ST & R &  5.2  & 34   &  0.36  & 3.4   &     4.30 &  V   & 11.99(33) & V    &(b) \\
             &    &         &     &   &       &      &        &       &          &      &           &      &    \\
J1058$+$0133 &  2 &25.03.07 &  ST & R &5.8    & 11   &  0.12  &  2.0  & 3.20     & V    & 58.56(10) & V    &(e)\\
             &    &16.04.07 &  ST & R &3.8    & 15   &  0.17  &  0.8  & 0.53     & N    & 1.37(14)  & N    &(e)\\
             &    &23.04.07 &  ST & R &4.4    & 10   &  0.23  &  1.8  & 2.57     & V    & 12.51(9)  & V    &(e)\\
             &    &         &     &   &       &      &        &       &          &      &           &      &    \\
J1159$+$2914 &  2 & 19.01.94&     & R &4.0    &      &  0.90  &   12.0&          & V    &           &      &(c) \\
             &    & 20.01.94&     & R &5.5    &      &  0.80  &    3.0&          & V    &           &      &(c) \\
             &    & 21.01.94&     & R &5.1    &      &  0.90  &   10.0&          & V    &           &      &(c) \\
             &    & 22.01.94&     & R &5.4    &      &  1.10  &   11.0&          & V    &           &      &(c) \\
             &    & 23.01.94&     & R &5.1    &      &  1.00  &   4.0 &          & V    &           &      &(c) \\
             &    & 24.01.94&     & R &4.0    &      &  1.30  &   7.0 &          & V    &           &      &(c) \\
             &    &         &     &   &       &      &        &       &          &      &           &      &    \\
J1217$+$3007 &   2&20.03.99 &  ST & R &  7.0  & 21   &        & 3.50  &     5.50 &  V   &           &     & (a)   \\
             &    &25.02.00 &  ST & R &  5.9  & 28   &        &       &          & N    &           &     & (a)   \\
             &    &31.03.00 &  ST & R &  5.0  & 27   &        &       &          & N    &           &     & (a)   \\
             &    &19.04.02 &  ST & R &  6.8  & 23   &        & 1.80  &    4.90  &  V   &           &     & (a)   \\
             &    &         &     &   &       &      &        &       &          &      &           &     &    \\
J1218$-$0119 &  2 &11.03.02 &  ST & R &8.0    & 22   &        &   7.3 & 3.20     & V    &           &      &(a) \\
             &    &13.03.02 &  ST & R &8.5    & 24   &        &   3.8 & 2.60     & V    &           &      &(a) \\
             &    &15.03.02 &  ST & R &3.9    & 11   &        &   5.5 & 3.50     & V    &           &      &(a) \\
             &    &16.03.02 &  ST & R &8.2    & 22   &        &  14.1 & $>$5.54  & V    &           &      &(a) \\
             &    &         &     &   &       &      &        &       &          &      &           &     &    \\
J1221$+$3010 &  1 &08.03.10 & IGO & R &  6.2  & 15   &  0.24  & 0.9   &     1.18 & N    & 1.96(14)  & N   &(b) \\
             &    &18.03.10 &  ST & R &  4.7  & 25   &  0.49  & 1.0   &     0.26 & N    & 2.12(24)  & N   &(b) \\
             &    &22.05.10 &  ST & R &  3.9  & 19   &  0.88  & 3.3   &     0.41 & N    & 2.23(18)  & PV  &(b) \\
             &    &         &     &   &       &      &        &       &          &      &           &     &    \\
J1221$+$2813 &  1 &19.03.04 & HCT & V &  5.4  & 74   &  0.32  & 5.2   &  $>$5.54 &  V   & 16.65(72) & V   &(b) \\
             &    &20.03.04 & HCT & V &  6.6  & 97   &  0.49  & 8.2   &  $>$5.54 &  V   & 62.42(96) & V   &(b) \\
             &    &18.03.05 &  ST & R &  4.0  & 26   &  0.33  & 2.0   &     1.74 &  N   & 3.21(25)  & V   &(b) \\
             &    &05.04.05 &  ST & R &  6.9  & 38   &  0.20  & 3.2   &     4.50 &   V  & 27.71(37) & V   &(b) \\
             &    &         &     &   &       &      &        &       &          &      &           &     &    \\
J1256$-$0547 &  1 &26.01.06 &  ST & R &  4.2  & 19   &  0.17  &  2.5  & $>$5.54  & V    &13.59(18)  & V   & (b)\\
             &    &28.02.06 &  ST & R &  6.1  & 40   &  0.25  &  10.2 & $>$5.54  & V    &403.89(39) & V   & (b)\\
             &    &20.04.09 &  ST & R &  4.9  & 20   &  0.44  &  22.0 & $>$5.54  & V    &1069.51(19)& V   & (b)\\
             &    &         &     &   &       &      &        &       &          &      &           &     &    \\
J1310$+$3220 & 2  &26.04.00 &  ST & R &5.6    & 16   &        &       &          & N    &           &     &(a) \\
             &    &17.03.02 &  ST & R &7.7    & 19   &        &   3.4 & 3.1      & V    &           &     &(a) \\
             &    &24.04.02 &  ST & R &5.8    & 14   &        &       &          & N    &           &     &(a) \\
             &    &02.05.02 &  ST & R &5.1    & 15   &        &       &          & N    &           &     &(a) \\
             &    &         &     &   &       &      &        &       &          &      &           &     &    \\
J1428$+$4240 &  1 &21.04.04 & HCT & V &  5.8  & 32   &  0.46  &  2.8  & 1.82     & N    & 2.90(31)  & V   &(b) \\
             &    &22.04.09 &  ST & R &  4.0  & 16   &  0.37  &  1.1  & 0.28     & N    & 1.99(15)  &  N  &(b) \\
             &    &29.04.09 &  ST & R &  6.4  & 27   &  0.73  &  2.1  & 0.65     & N    & 1.42(26)  &  N  &(b) \\
             &    &         &     &   &       &      &        &       &          &      &           &     &    \\
J1512$-$0906 &  2 &28.04.98 &     & V &3.8    &      &  0.30  &       &          & N    &           &     &(i) \\
             &    &29.04.98 &     & V &4.0    &      &  0.40  &       &          & N    &           &     &(i) \\
             &    &30.04.98 &     & V &4.0    &      &  0.80  &   4.2 & 1.45     & N    &           &     &(d) \\
             &    &06.06.99 &     & V &7.2    &      &  0.90  &   3.4 & 1.05     & N    &           &     &(d) \\
             &    &07.06.99 &     & V &7.3    &      &  0.90  &   4.2 & 1.27     & N    &           &     &(d) \\
             &    &14.06.05 &  ST & R &4.0    & 09   &  0.12  &   1.6 & 1.53     & N    & 12.02(8)  & V   &(e) \\
             &    &01.05.09 &  ST & R &5.6    & 22   &  0.31  &   4.7 & 2.63     & V    & 29.85(21) & V   &(e) \\
             &    &20.05.09 &  ST & R &4.8    & 23   &  0.43  &   3.1 & 0.98     & N    & 4.20(22)  & V   &(e) \\
             &    &         &     &   &       &      &        &       &          &      &           &     &    \\
J1555$+$1111 &  1 &05.05.99 &  ST & R &  3.6  & 20   &        & 2.3   & $>$6.60  & V    &           &     & (i) \\
             &    &06.06.99 &  ST & R &  7.1  & 40   &        &       &          & N    &           &     & (i) \\
             &    &24.06.09 &  ST & R &  3.8  & 23   &  0.17  & 4.2   & $>$5.54  & V    & 50.81(22) &  V  & (b) \\
             &    &15.05.10 &  ST & R &  6.1  & 25   &  0.29  & 2.8   & $>$5.54  & V    & 12.99(24) &  V  & (b) \\
             &    &16.05.10 &  ST & R &  6.1  & 31   &  0.30  & 2.3   &    4.63  & V    & 6.33(30)  &  V  & (b) \\
             &    &         &     &   &       &      &        &       &          &      &           &     &    \\
J2225$-$0457 &  2 &29.09.88 &     & R &4.8    &      &   0.80 &   9.0 &          & V    &           &     &(c) \\
             &    &01.10.88 &     & R &6.7    &      &   0.90 &       &          & N    &           &     &(c) \\
             &    &08.10.10 &     & R &5.1    & 16   &   0.64 &   6.8 &   1.12   & N    & 14.35(15) & V   & (e) \\
             &    &         &     &   &       &      &        &       &          &      &           &     &     \\
J2253$+$1608 &  2 &12.09.90 &     & R &6.9    &      &   1.00 &   6.0 &          & V    &           &     &(c) \\
             &    &13.09.90 &     & R &7.5    &      &   1.10 &   4.0 &          & V    &           &     &(c) \\
\enddata

{
$^\dag$ V= variable, N = non-variable, PV = probable variable \\
$^\ddag$ DOF = degrees of freedom \\
$^\pounds$Reference for INOV data: (a) Sagar et al.\ (2004); (b) Present work; (c) Noble (1995); (d) Romero et al.\ (2002); (e) Goyal et al. (2011)
  (f) Pollock, Webb \& Azarnia (2007); (g) Gupta et al.\ (2008); (h) Goyal et al.\ (2009); (i) Stalin et al.\ (2005)
}
\end{deluxetable}

\newpage
\begin{figure*}
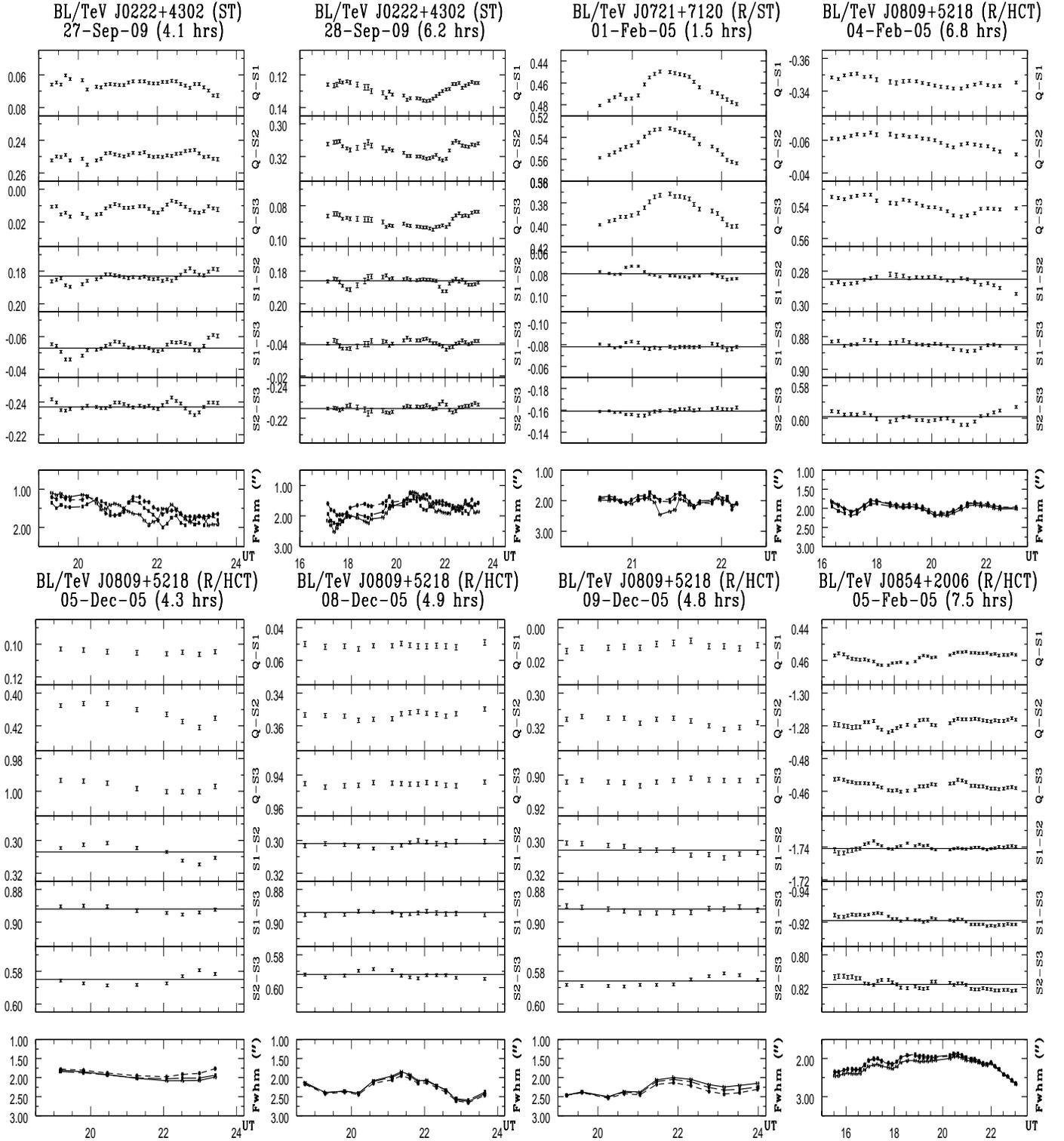

\hspace*{-1.0cm}
\hbox{
\includegraphics[height=10.0cm,width=04.5cm]{fig_J0222+4302_27sep09_ST.epsi}
\includegraphics[height=10.0cm,width=04.5cm]{fig_J0222+4302_28sep09_ST.epsi}
\includegraphics[height=10.0cm,width=04.5cm]{fig_J0721+7120_01feb05_ST.epsi}
\includegraphics[height=10.0cm,width=04.5cm]{fig_J0809+5218_R_HCT_04feb05.epsi}
}
\vspace*{0cm}
\hspace*{-1.0cm}
\hbox{
\includegraphics[height=10.0cm,width=04.5cm]{fig_J0809+5218_R_HCT_05dec05.epsi}
\includegraphics[height=10.0cm,width=04.5cm]{fig_J0809+5218_R_HCT_08dec05.epsi}
\includegraphics[height=10.0cm,width=04.5cm]{fig_J0809+5218_R_HCT_09dec05.epsi}
\includegraphics[height=10.0cm,width=04.5cm]{fig_J0854+2006_05feb05_HCT.epsi}
}
\caption{ Intranight DLCs of the 9 TeV blazars (Set 1) and J0854+2006 (Set 2) monitored in the present study. 
The source name, date of monitoring, its duration,
the filter and the telescope used are mentioned at the top of each frame.
For each night the bottom panel shows the variation of the atmospheric seeing
through the monitoring duration. 
}. \medskip
\label{fig:1}
\end{figure*}
\clearpage

\begin{figure*}
\hspace*{-1.0cm}
\hbox{
\includegraphics[height=10.0cm,width=04.5cm]{fig_J0854+2006_12apr05_ST.epsi}
\includegraphics[height=10.0cm,width=04.5cm]{fig_J1015+4926_06feb10_ST.epsi}
\includegraphics[height=10.0cm,width=04.5cm]{fig_J1015+4926_19feb10_ST.epsi}
\includegraphics[height=10.0cm,width=04.5cm]{fig_J1015+4926_07mar10_ST.epsi}
}
\vspace*{0cm}
\hspace*{-1.0cm}
\hbox{
\includegraphics[height=10.0cm,width=04.5cm]{fig_J1221+3010_08mar10_IGO.epsi}
\includegraphics[height=10.0cm,width=04.5cm]{fig_J1221+3010_18mar10_ST.epsi}
\includegraphics[height=10.0cm,width=04.5cm]{fig_J1221+3010_22may10_ST.epsi}
\includegraphics[height=10.0cm,width=04.5cm]{fig_J1221+2813_19mar04_HCT.epsi}
}
\begin{center}
{{\bf Figure~\ref{fig:1}}. \textit {continued}}
\end{center}
\end{figure*}
\clearpage
\newpage

\begin{figure*}
\hspace*{-1.0cm}
\hbox{
\includegraphics[height=10.0cm,width=04.5cm]{fig_J1221+2813_20mar04_HCT.epsi}
\includegraphics[height=10.0cm,width=04.5cm]{fig_J1221+2813_18mar05_ST.epsi}
\includegraphics[height=10.0cm,width=04.5cm]{fig_J1221+2813_05apr05_ST.epsi}
\includegraphics[height=10.0cm,width=04.5cm]{fig_J1256-0547_26jan06_ST_tev.epsi}
}
\vspace*{0cm}
\hspace*{-1.0cm}
\hbox{
\includegraphics[height=10.0cm,width=04.5cm]{fig_J1256-0547_28feb06_ST_tev.epsi}
\includegraphics[height=10.0cm,width=04.5cm]{fig_J1256-0547_20apr09_ST_tev.epsi}
\includegraphics[height=10.0cm,width=04.5cm]{fig_J1428+4240_V_HCT_21apr04.epsi}
\includegraphics[height=10.0cm,width=04.5cm]{fig_J1428+4240_R_ST_22apr09.epsi}
}
\begin{center}
{{\bf Figure~\ref{fig:1}}. \textit {continued}}
\end{center}
\end{figure*}
\clearpage
\newpage

\begin{figure*}
\hspace*{-1.0cm}
\hbox{
\includegraphics[height=10.0cm,width=04.5cm]{fig_J1428+4240_R_ST_29apr09.epsi}
\includegraphics[height=10.0cm,width=04.5cm]{fig_J1555+1111_24jun09_ST.epsi}
\includegraphics[height=10.0cm,width=04.5cm]{fig_J1555+1111_15may10_ST.epsi}
\includegraphics[height=10.0cm,width=04.5cm]{fig_J1555+1111_16may10_ST.epsi}
}
\begin{center}
{{\bf Figure~\ref{fig:1}}. \textit {continued}}
\end{center}
\end{figure*}
\clearpage

\newpage
\begin{figure*}
\includegraphics[height=22.0cm,width=16.0cm]{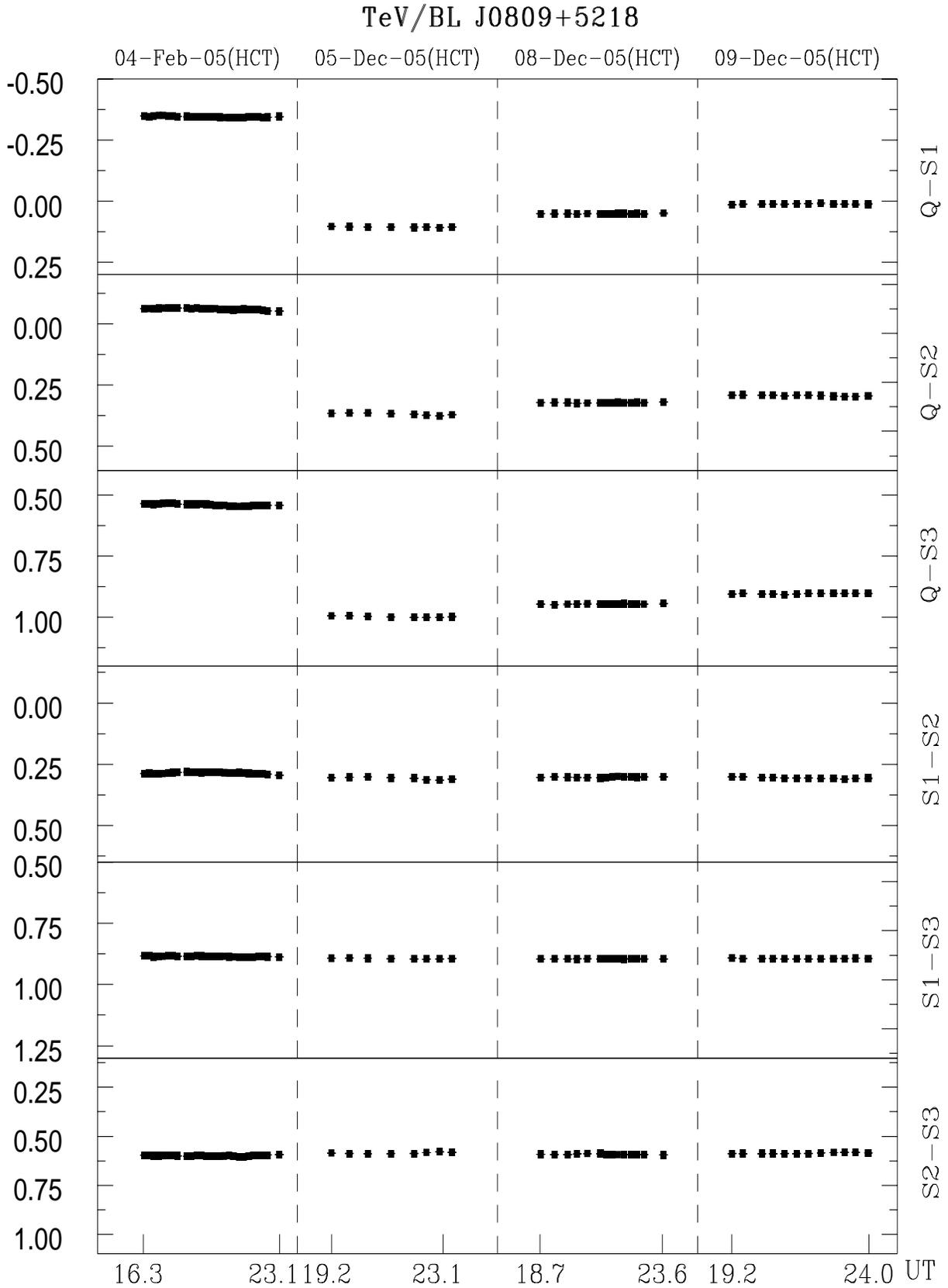}
\caption{Long-Term Optical Variability (LTOV) DLCs of the TeV blazars in Set 1 (see text).} 
\label{fig:3}
\end{figure*}
\clearpage

\newpage
\begin{figure*}
\includegraphics[height=22.0cm,width=16.0cm]{ltov_J1015.epsi}
\begin{center}
{{\bf Figure~\ref{fig:3}}. \textit {continued}}
\end{center}
\end{figure*}
\clearpage

\newpage
\begin{figure*}
\includegraphics[height=22.0cm,width=16.0cm]{ltov_J1221.epsi}
\begin{center}
{{\bf Figure~\ref{fig:3}}. \textit {continued}}
\end{center}
\end{figure*}
\clearpage

\newpage
\begin{figure*}
\includegraphics[height=22.0cm,width=16.0cm]{ltov_hpJ1256.epsi}
\begin{center}
{{\bf Figure~\ref{fig:3}}. \textit {continued}}
\end{center}
\end{figure*}
\clearpage

\newpage
\begin{figure*}
\includegraphics[height=22.0cm,width=16.0cm]{ltov_J1556.epsi}
\begin{center}
{{\bf Figure~\ref{fig:3}}. \textit {continued}}
\end{center}
\end{figure*}
\clearpage

\newpage
\begin{figure*}
\includegraphics[height=15.0cm,width=15.0cm]{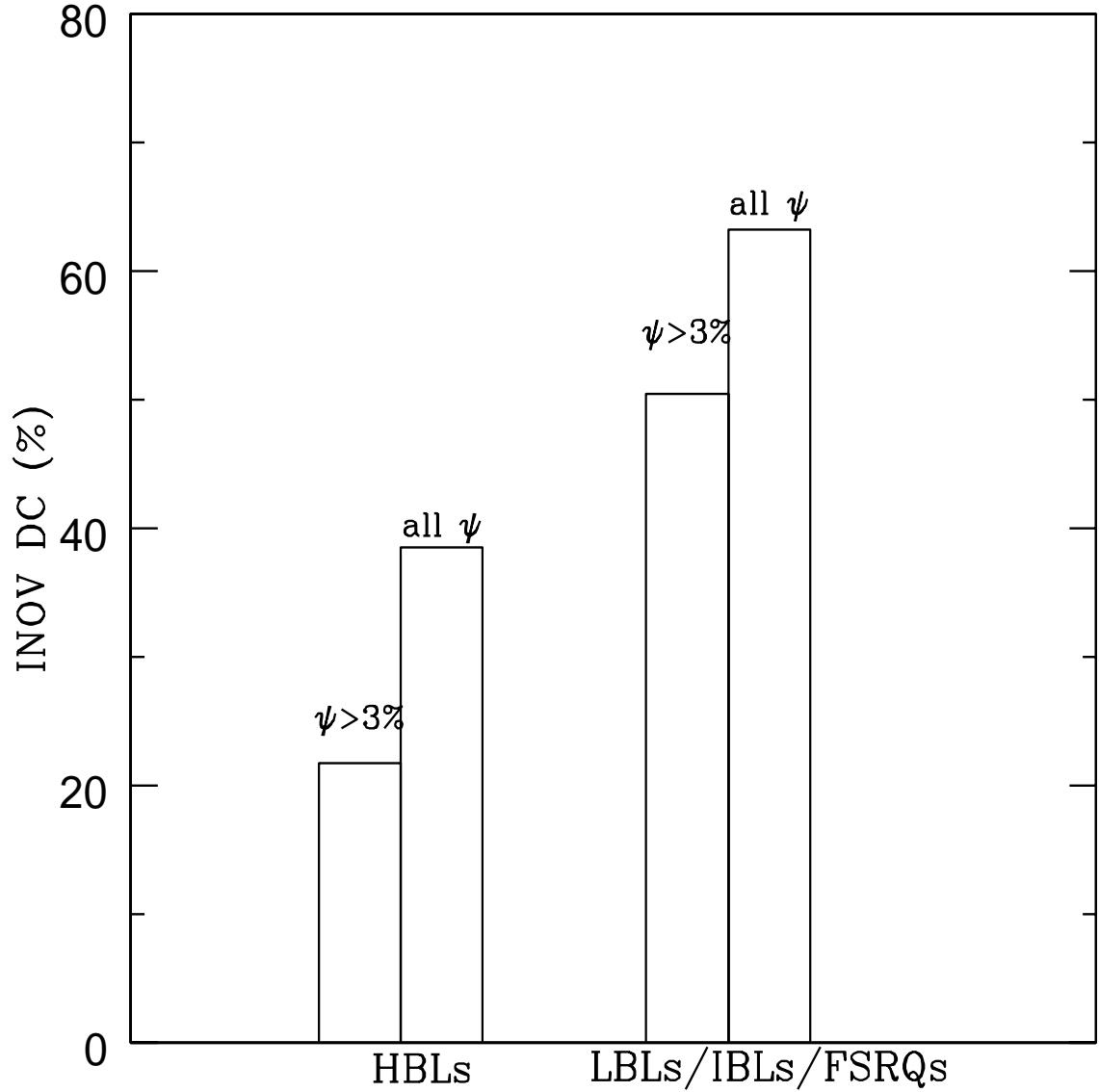}
\caption{Histogram of the INOV duty cycle (present work) derived for the different blazar classes, for two ranges of INOV fractional amplitude $\psi$.} 
\label{fig:4}
\end{figure*}
\clearpage


\newpage
\begin{thebibliography}{}

  \bibitem[\protect\citeauthoryear{Abdo et al}%
   {2009}]{} Abdo A. A., et al., 
   2009, ApJ, 700, 597 
  \bibitem[\protect\citeauthoryear{Abdo et al}%
   {2010a}]{} Abdo A. A., et al., 
   2010a, ApJ, 716, 30 
  \bibitem[\protect\citeauthoryear{Abdo et al}%
   {2010b}]{} Abdo A. A., et al., 
   2010b, ApJS, 188, 405
  \bibitem[\protect\citeauthoryear{Aharonian et al}%
   {2002}]{} Aharonian F., et al.,
   2002, A\&A, 384, 23 
  \bibitem[\protect\citeauthoryear{Agudo et al}%
   {2010}]{} Agudo I., et al., 2010, preprint (astroph/1011.6454v1)  
  \bibitem[\protect\citeauthoryear{Ammando et al}%
   {2006}]{} Ammando F. D., 2010, in SciNeGHE 2010 Workshop, September 8-10, 2010, Trieste, Italy. To appear in Il Nuovo Cimento C - Colloquia on physics.
   preprint (astroph/1012.1120)
  \bibitem[\protect\citeauthoryear{Andruchow et al}%
   {2005}]{} Andruchow I., Romero G. E., Cellone S., 2005, A\&A, 442, 97 
  \bibitem[\protect\citeauthoryear{Attridge et al}%
   {1999}]{} Attridge, J. M., Roberts, D. H., Wardle, J. F. C., 1999, ApJ, 518, L87
   \bibitem[\protect\citeauthoryear{Atwood et al.}%
   {2009}]{} Atwood W. B., 2009, ApJ, 697, 1071
   \bibitem[\protect\citeauthoryear{Begelman et al}%
   {1979}]{} Begelman M. C., Blandford R. D., Rees M. J., 1984, RvMP, 56, 255 
  \bibitem[\protect\citeauthoryear{Begelman et al}%
   {2008}]{} Begelman M. C., Fabian A. C., Rees M. J., 2008, MNRAS, 384, L19 
  \bibitem[\protect\citeauthoryear{Bramel et al}%
   {2005}]{} Bramel D. A., et al.,
   2005, ApJ, 629, 108  
  \bibitem[\protect\citeauthoryear{Britzen et al}%
   {2008}]{} Britzen S., et al.,
   2008, A\&A, 484, 119
  \bibitem[\protect\citeauthoryear{Britzen et al}%
   {2010}]{} Britzen S., et al.,
   2010, A\&A, 515, 105  
  \bibitem[\protect\citeauthoryear{Carini et al}%
   {1990}]{} Carini M. T., Miller H. R., Goodrich B. D., 1990, AJ, 100, 347
  \bibitem[\protect\citeauthoryear{Cellone et al}%
   {2000}]{} Cellone S. A., Romero G. E., Combi J. A., 2000, AJ, 119, 1534
  \bibitem[\protect\citeauthoryear{Cellone et al}%
   {2007}]{} Cellone S. A., Romero G. E., Araudo A. T., 2007, MNRAS, 374, 357
  \bibitem[\protect\citeauthoryear{Ciprini et al}%
   {2007}]{} Ciprini S., et al, 2007, A\&A, 467, 465
  \bibitem[\protect\citeauthoryear{Chen et al}%
   {2006}]{} Chen Y., Gu M., Shen Z.-Q., Fan Z., 2006, MNRAS, 370, 1885 
  \bibitem[\protect\citeauthoryear{Costamante et al}%
   {2001}]{} Costamante L., et al.,
   2001, A\&A, 371, 512
  \bibitem[\protect\citeauthoryear{Costamante et al}%
   {2009}]{} Costamante L., Aharonian F., Buehler R., Khangulyan D., Reimer A., Reimer O., 2009, preprint (astroph/0907.3966) 
  \bibitem[\protect\citeauthoryear{Czerny et al}%
   {2020}]{} Czerny B., Hryniewicz K., preprint (astroph/1010.6201)
  \bibitem[\protect\citeauthoryear{Dai et al.}%
   {2001}]{} Dai B. Z., Xie G. Z., Li K. H., Zhou S. B., Liu W. W., Jiang Z. J., 2001, AJ, 122, 2901
  \bibitem[\protect\citeauthoryear{de Diego}%
   {2010}]{} de Diego J. A., 2010, AJ, 139, 1269
  \bibitem[\protect\citeauthoryear{Edelson et al.}%
   {2002}]{}Edelson R., Turner T. J., Pounds K., Vaughan S., Markowitz A., Marshall H., Dobbie P., Warwick R., 2002, ApJ, 568, 610 
  \bibitem[\protect\citeauthoryear{Fabian \& Rees}%
   {1979}]{} Fabian A. C., Rees M. J., 1979, MNRAS, 187   
  \bibitem[\protect\citeauthoryear{Fan et al}%
   {1997}]{} Fan J. H., Cheng K. S., Zhang L., Liu C. H., 1997, A\&A, 327, 947   
  \bibitem[\protect\citeauthoryear{Giannios et al}%
   {2009}]{} Giannios D., Uzdensky D. A., Begelman M. C., 2009, MNRAS, 395, 29  
  \bibitem[\protect\citeauthoryear{Ghisellini et al}%
   {2004}]{} Ghisellini G., Haardt F., Matt G, 2004, A\&A, 413, 523 
   \bibitem[\protect\citeauthoryear{Ghisellini et al}%
   {1989}]{} Ghisellini G., Maraschi, L., 1989, ApJ, 340, 181
    \bibitem[\protect\citeauthoryear{Ghisellini et al}%
   {1997}]{} Ghisellini G., et al., 1997, A\&A, 327, 61  
    \bibitem[\protect\citeauthoryear{Ghisellini et al}%
   {2009}]{} Ghisellini G., Maraschi, L., Tavecchio, F. 2009, MNRAS, 396, L105
  \bibitem[\protect\citeauthoryear{Giroletti et al}%
   {2004}]{} Giroletti M., Giovannini G., Taylor G. B., Falomo R., 2004, ApJ, 613, 752 
  \bibitem[\protect\citeauthoryear{Giroletti et al}%
   {2006}]{} Giroletti M., Giovannini G., Taylor G. B., Falomo R., 2006, ApJ, 646, 801 
 \bibitem[\protect\citeauthoryear{GK et al}%
   {1984}]{}  Gopal-Krishna, Singal, A.K., Krishnamohan, S., 1984, A\&A, 140, L19   
  \bibitem[\protect\citeauthoryear{GK et al}%
   {1991}]{}  Gopal-Krishna, Subramanian K., 1991, Nature, 349, 766
  \bibitem[\protect\citeauthoryear{GK et al}%
   {2003}]{}  Gopal-Krishna, Stalin C. S., Sagar R.,
    Wiita P. J., 2003, ApJ, 586, L25
  \bibitem[\protect\citeauthoryear{GK et al}%
   {2004}]{} Gopal-Krishna, Dhurde, S., Wiita P. J., 2004, ApJL, 615, L81 
  \bibitem[\protect\citeauthoryear{GK et al}%
   {2006}]{} Gopal-Krishna, Wiita P. J., Dhurde S., 2006, MNRAS, 369, 1281 
  \bibitem[\protect\citeauthoryear{GK et al}%
   {2007}]{} Gopal-Krishna, Dhurde S., Sirkar P., Wiita P. J., 2007, MNRAS, 377, 446 
  \bibitem[\protect\citeauthoryear{Goyal et al}%
   {2009}]{} Goyal A., et al.,
   2009, MNRAS, 399, 1622
  \bibitem[\protect\citeauthoryear{Goyal et al}%
   {2011}]{} Goyal A., Gopal-krishna, Wiita P. J. W., Anupama G. C., Sahu D. K. S., Sagar R., 2011,  submitted
  \bibitem[\protect\citeauthoryear{Guilbert et al}%
   {1983}]{} Guilbert P. W., Fabian A. C., Rees M. J., 1983, MNRAS, 205, 593
  \bibitem[\protect\citeauthoryear{Gupta et al}%
   {2008}]{} Gupta A. C., Fan J. H., Bai J. M., Wagner S. J., 2008, AJ, 135, 1384
  \bibitem[\protect\citeauthoryear{Hartman et al}%
   {1992}]{} Hartman R. C., et al.,
    1992, ApJL, 385, 1 
   \bibitem[\protect\citeauthoryear{Heidt \& Wagner}%
   {1998}]{} Heidt J., Wagner S. J., 1998, A\&A, 329, 853
  \bibitem[\protect\citeauthoryear{Helmboldt et al}%
   {2007}]{} Helmboldt J. F., et al., 2007, ApJ, 658, 203 
  \bibitem[\protect\citeauthoryear{Impiombato et al}%
   {2011}]{} Impiombato D., et al., 2011, ApJS, 192, 12 
  \bibitem[\protect\citeauthoryear{Jannuci et al}%
   {1993}]{} Jannuzi B. T., Smith P. S., Elston R., 1993, ApJS, 85, 265
  \bibitem[\protect\citeauthoryear{Jorstad et al}%
   {2001}]{} Jorstad S. G., Marscher A. P., Mattox J. R., Wehrle A. E., Bloom S. D., Yurchenko A. V., 2001, ApJS, 134, 181
  \bibitem[\protect\citeauthoryear{Kataoka et al}%
   {2001}]{} Kataoka J., et al.,
    2001, ApJ, 560, 659 
  \bibitem[\protect\citeauthoryear{Kellermann1969}%
   {2004}]{} Kellermann K. I., Pauliny-Toth I. I. K., 1969, ApJ, 155, 71 
  \bibitem[\protect\citeauthoryear{Kellermann et al}%
   {2004}]{} Kellermann K. I., et al., 
    2004, ApJ, 609, 539 
  \bibitem[\protect\citeauthoryear{Kharb et al}%
   {2008}]{} Kharb P., Gabuzda D., Shastri P., 2008, MNRAS, 384, 230 
  \bibitem[\protect\citeauthoryear{Kidger et al}%
   {1990}]{} Kidger M. R., deDiego, J. A., 1990, A\&A, 227, L25 
  \bibitem[\protect\citeauthoryear{Kovalev et al}%
   {1999}]{} Kovalev Y. Y., Nizhelsky N. A., Kovalev Y. A., Berlin A. B., Zhekanis G. V., Mingaliev M. G., Bogdantsov A. V., 1999, A\&AS, 139, 545 
  \bibitem[\protect\citeauthoryear{Kovalev et al}%
   {2005}]{} Kovalev Y. Y., et al., 
   2005 AJ, 130, 2473
  \bibitem[\protect\citeauthoryear{Kovalev et al}%
   {2009}]{} Kovalev Y. Y, et al.,
   2009 ApJL, 696, 17   
   \bibitem[\protect\citeauthoryear{Krawczynski et al}%
   {2002}]{} Krawczynski H., Coppi P. S., Aharonian F., 2002, MNRAS, 336, 721 
  \bibitem[\protect\citeauthoryear{Krongold et al}%
   {2010}]{} Krongold Y., Binette L., Hernandez-Ibarra F., 2010, ApJ, 724 L203
  \bibitem[\protect\citeauthoryear{Kundt et al}%
   {1980}]{} Kundt W., Gopal-Krishna, 1980, Nature, 288, 149
  \bibitem[\protect\citeauthoryear{Kundt et al}%
   {2004}]{} Kundt W., Gopal-Krishna, 2004, JApA, 25, 115 
  \bibitem[\protect\citeauthoryear{Li et al}%
   {2010}]{} Li H. Z., Xie G. Z., Yi T. F., Chen L. E., Dai H., 2010, ApJ, 709, 1407
   \bibitem[\protect\citeauthoryear{Lister et al}%
   {2005}]{} Lister M. L., Homan D. C., 2005, ApJL, 130, 1389    
   \bibitem[\protect\citeauthoryear{Lister et al}%
   {2009a}]{} Lister M. L., Homan D. C., Kedler M., Kellermann K. I., Kovalev Y. Y., Ros E., Savolainen T., Zensus A., 2009a, ApJL, 696, 22      
   \bibitem[\protect\citeauthoryear{Lister et al}%
   {2009b}]{} Lister M. L., et al., 
   2009b, AJ, 138, 1874    
   \bibitem[\protect\citeauthoryear{Lobanov}%
   {2010}]{} Lobanov A.P., in ``Steady jets and transient jets'',   
    Bonn, 7-8 April 2010. published in Memorie della Societa Astronomica Italiana, v.81, n.4 (2010)
  \bibitem[\protect\citeauthoryear{MAGIC}%
   {2008}]{} MAGIC collaboration; Albert J., et al., 
    2008, Science, 320, 1752 
   \bibitem[\protect\citeauthoryear{Marscher}%
   {1996}]{} Marscher A. P., 1996, in Blazar continuum variability, eds. H.R.\ Miller,
   J.R.\ Webb, and J.C. Noble, ASP Conf.\ Ser.\ Vol. 110, (San Francisco: ASP), p.\ 248
   \bibitem[\protect\citeauthoryear{Marscher et al}%
   {2002}]{} Marscher A. P., Jorstad S. G., Mattox J. R., Wehrle A. E., 2002, ApJ, 577, 85      
   \bibitem[\protect\citeauthoryear{Marscher et al}
   {2008}]{} Marscher, A.P. et al. 2008, Nature, 452, 966 
  \bibitem[\protect\citeauthoryear{Mihov et al}%
   {2008}]{} Mihov B., Bachev R., Slavcheva-Mihova L., Strigachev A., Semkov E., Petrov G., 2008, AN, 329, 77
  \bibitem[\protect\citeauthoryear{Monet et al}%
   {2003}]{} Monet D. G., et al.,
    2003, AJ, 125, 984 
  \bibitem[\protect\citeauthoryear{ Noble}%
   {1995}]{} Noble J. C., 1995, PhD thesis, Georgia State University
  \bibitem[\protect\citeauthoryear{Perri et al}%
   {2003}]{} Perri M., et al.,
   2003, 407, 453 
  \bibitem[\protect\citeauthoryear{Petrucci}%
   {2010}]{} Petrucci P. O., Boutelier T., Henri G., 2010, preprint (astroph/1010.5895)
  \bibitem[\protect\citeauthoryear{Piner et al}%
   {2004}]{} Piner B. G., Edwards P. G., 2004, ApJ, 600, 115
  \bibitem[\protect\citeauthoryear{Piner et al}%
   {2008}]{} Piner B. G., Pant N., Edwards P. G., 2008, ApJ, 678, 64
  \bibitem[\protect\citeauthoryear{Pollock et al}%
   {2007}]{} Pollock J. T., Webb J. R., Azarnia G., 2007, ApJ, 133, 487
   \bibitem[\protect\citeauthoryear{Rani et al}%
   {2010}]{} Rani, B., Gupta, A.C., Joshi, U.C., Ganesh, S., Wiita, P.J., 2010, ApJ, 719, L153
  \bibitem[\protect\citeauthoryear{Rector et al}%
   {2003}]{} Rector T. A., Gabuzda D. C., Stocke J. T., 2003, AJ, 125, 1060  
  \bibitem[\protect\citeauthoryear{Romero et al}%
   {1999}]{} Romero G. E., Cellone S. A., Combi J. A., 1999, A\&AS, 135, 477
   \bibitem[\protect\citeauthoryear{Romero et al}%
   {2002}]{} Romero G. E., Cellone, S. A., Combi J. A., Andruchow I., 2002, A\&A, 390, 431
   \bibitem[\protect\citeauthoryear{Sagar}%
   {1999}]{} Sagar R., 1999, Curr. Sci, 77, 643
  \bibitem[\protect\citeauthoryear{Sagar et al}%
   {1999}]{} Sagar R., Gopal-Krishna, Mohan V., Pandey A. K., Bhatt B. C., Wagner S.J., 1999, A\&AS, 134, 453
  \bibitem[\protect\citeauthoryear{Sagar et al}%
   {2004}]{} Sagar R., Stalin C. S., Gopal-Krishna, Wiita P. J.,
    2004, MNRAS, 348, 176
   \bibitem[\protect\citeauthoryear{Sambruna et al.}%
   {1996}]{} Sambruna R. M., Maraschi L., Urry C. M. 1996, ApJ, 463, 444 
  \bibitem[\protect\citeauthoryear{Savolainen et al}%
   {2010}]{} Savolainen T., Homan D. C., Hovatta T., Kadler M., Kovalev Y. Y., Lister M. L., Ros E., Zensus J. A., 2010, A\&A, 512, 24
   \bibitem[\protect\citeauthoryear{Schlegel et al}%
   {1998}]{} Schlegel D. J., Finkbeiner D. P., Davis M., 1998, ApJ, 500, 525
  \bibitem[\protect\citeauthoryear{Schlickeiser}%
   {1996}]{} Schlickeiser R., 1996, A\&AS, 120, 481
  \bibitem[\protect\citeauthoryear{Spergel et al}%
   {2007}]{} Spergel D. N., 2007, ApJS, 170, 377
   \bibitem[\protect\citeauthoryear{Stalin et al}%
   {2004a}]{} Stalin C. S., Gopal-Krishna,
   Sagar R., Wiita P. J., 2004a, JApA, 25, 1
  \bibitem[\protect\citeauthoryear{Stalin et al}%
   {2004b}]{} Stalin, C. S., Gopal-Krishna, Sagar, R., Wiita, P. J., 2004b, MNRAS, 350, 175
  \bibitem[\protect\citeauthoryear{Stalin et al}%
   {2005}]{} Stalin C. S., Gupta A. C., Gopal-Krishna, Wiita P. J.,
   Sagar R., 2005, MNRAS, 356, 607
   \bibitem[\protect\citeauthoryear{Stetson}%
    {1987}]{} Stetson P. B., 1987, PASP, 99, 191
  \bibitem[\protect\citeauthoryear{Stocke et al84}%
    {1984}]{} Stocke J. T., Foltz C. B.,  Weymann R. J., Christiansen W. A., 1984, ApJ, 280, 476
  \bibitem[\protect\citeauthoryear{Stockman et al}%
   {1984}]{} Stockman H. S., Moore R. L., Angel J. R. P., 1984, ApJ, 279, 485
  \bibitem[\protect\citeauthoryear{Takalo et al}%
    {1994}]{} Takalo L. O., Sillanp{\"a}{\"a} A., Nilsson K., 1994, A\&AS, 107, 497
  \bibitem[\protect\citeauthoryear{Tanihata et al}%
    {1992}]{} Stocke J. T., Morris S. L., Weymann R. J., Foltz C. B., 1992, ApJ, 396, 487
  \bibitem[\protect\citeauthoryear{Tanihata et al}%
   {2001}]{} Tanihata C., Urry C. M., Takahashi T., Kataoka J., Wagner S. J., Madejski G. M., 
   Tashiro M., Kouda M., 2001, ApJ, 563, 569
  \bibitem[\protect\citeauthoryear{Taylor et al}%
   {2007}]{} Taylor G. B., et al., 
   2007 ApJ, 671, 1355
  \bibitem[\protect\citeauthoryear{Treves et al}%
   {2007}]{} Treves A., Falomo R., Uslenghi M., 2007, A\&A, 473, 17
  \bibitem[\protect\citeauthoryear{Urry \& Padovani}%
   {1995}]{} Urry C. M., Padovani P., 1995, PASP, 107, 803
  \bibitem[\protect\citeauthoryear{Veron \& Veron}%
   {2006}]{} V\'eron-Cetty M.-P.; V\'eron P., 2006, A\&A, 455, 773
  \bibitem[\protect\citeauthoryear{Visvanathan \& Wills}%
   {1998}]{} Visvanathan N., Wills B. J., 1998, AJ, 116, 2119
  \bibitem[\protect\citeauthoryear{Wagner \& Witzel}%
   {1995}]{} Wagner S. J., Witzel A., 1995, ARA\&A, 33, 163
  \bibitem[\protect\citeauthoryear{Wagner et al}%
   {1996}]{} Wagner S. J., et al.,
   1996, AJ, 111, 2187 
  \bibitem[\protect\citeauthoryear{Webb \& Malkan}%
  {2000}]{}  Webb W., Malkan M., 2000, ApJ, 540, 652   
  \bibitem[\protect\citeauthoryear{Weekes}%
   {2008}]{} Weekes T. C., 2008, in High Energy Gamma-Ray Astronomy, Ed. F. A. Aharonian, W. Hofmann \& F. Rieger, AIP Conf.\ Ser., 1085,  3  
   \bibitem[\protect\citeauthoryear{Wehrle et al}%
   {1984}]{} Wehrle A. E., Morabito D. D., Preston R. A., 1984, ApJ, 89, 336
  \bibitem[\protect\citeauthoryear{Wiita}%
   {2006}]{} Wiita P. J., 2006, in Blazar Variability Workshop II: Entering the GLAST Era ASP Conference Series, Vol.\ 350,
   eds. H. R. Miller, K. Marshall, J. R. Webb, and M. F. Aller. (San Francisco: ASP), p.\ 183
  \bibitem[\protect\citeauthoryear{Wills et al}
   {1992}]{} Wills B. J., Wills D., Breger M., Antonucci R. R. J., Barvianis R., 1992, ApJ, 398, 454
  \bibitem[\protect\citeauthoryear{Xie et al}%
   {2001}]{} Xie G. Z., Li K. H., Bai J. M., Dai B. Z., Liu W. W., Zhang X., Xing S. Y., 2001, ApJ, 548, 200



\end{thebibliography}
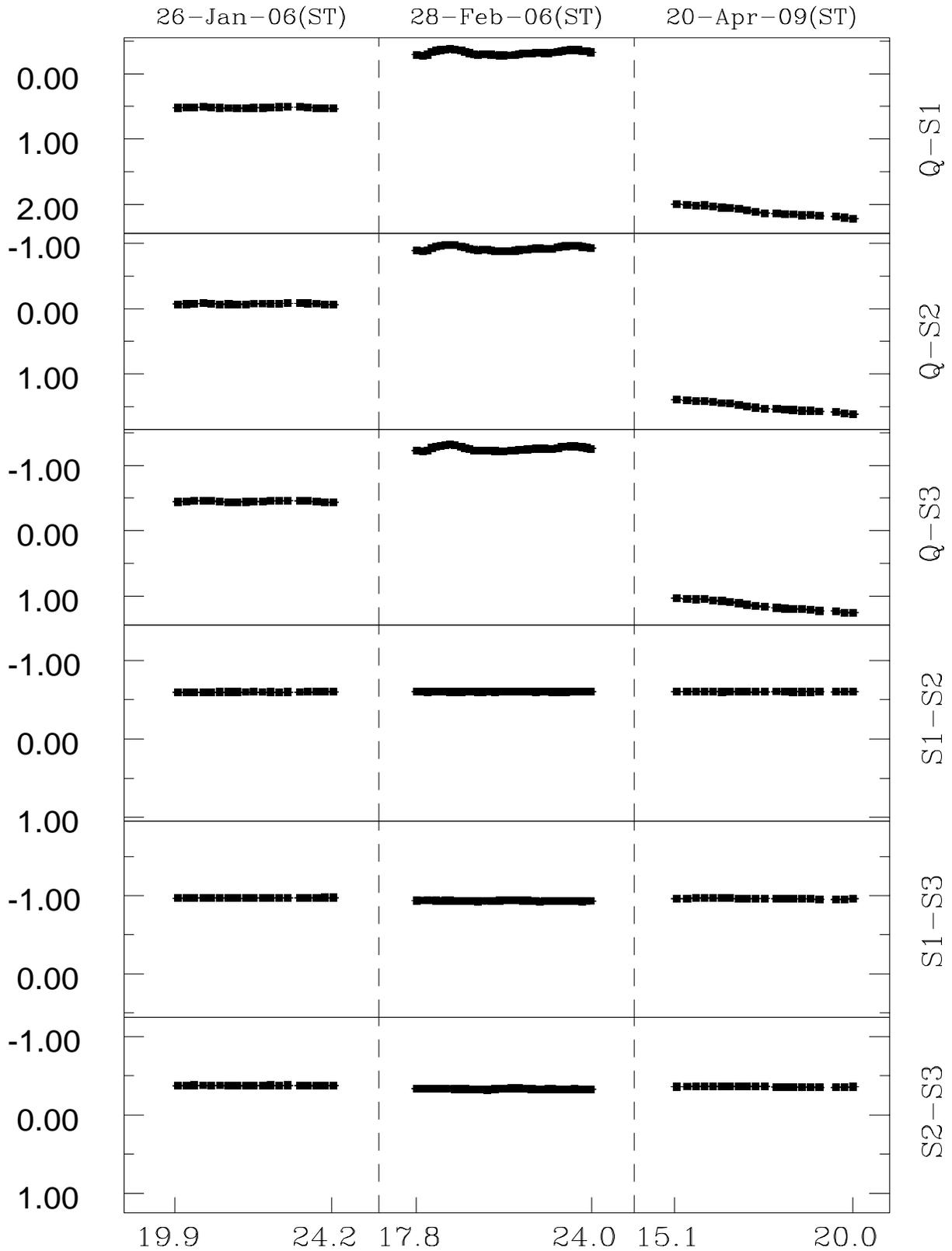
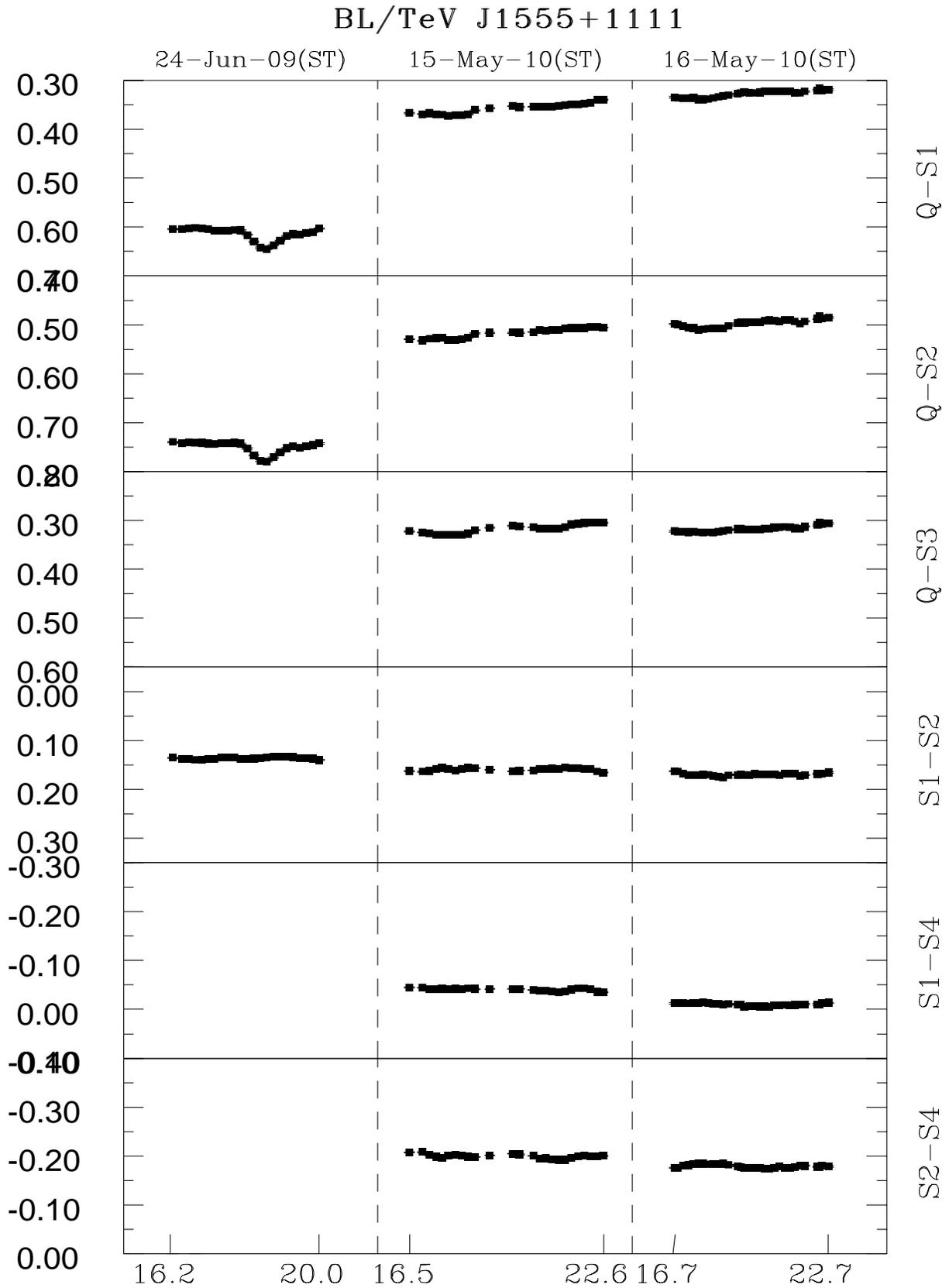
\end{document}